\documentclass[11pt]{article} 
\pdfoutput=1
\usepackage{geometry}                
\usepackage[bookmarksopen=false]{hyperref}                 
\usepackage{graphicx}
\usepackage{amssymb}
\usepackage{amsmath, amsfonts, multirow}
\usepackage{amsmath}
\usepackage{tikz}
\usepackage{pdflscape}
\usepackage{cite}

\def\eqn#1{\begin{equation}\begin{split}#1\end{split}\end{equation}}

\def\s2{\sqrt{2}}
\def\PP{{\mathbb P}}
\def\ZZ{{\mathbb Z}}
\def\CC{{\mathbb C}}

\title{D-brane Moduli Spaces and Superpotentials in a Two-Parameter Model}
\author{
Marco Baumgartl,\footnote{\tt marco.baumgartl@desy.de}\\[6pt]
{\it II.~Institut f\"ur Theoretische Physik, Universit\"at Hamburg}\\
{\it Luruper Chaussee 149, 22761 Hamburg, Germany}\\[12pt]
Ilka Brunner,\footnote{\tt ilka.brunner@physik.uni-muenchen.de} and Daniel Plencner,\footnote{\tt daniel.plencner@physik.uni-muenchen.de}\\[6pt]
{\it Arnold Sommerfeld Center f\"ur Theoretische Physik, LMU M\"unchen}\\
{\it Theresienstra{\ss}e 37, 80333 M\"unchen, Germany}\\
{\it Excellence Cluster Universe}\\
{\it Boltzmannstra{\ss}e 2, 85748 Garching, Germany}
}
\date{}                                           

\begin{document}
\maketitle

\vspace{-10cm}
\hfill {\small LMU-ASC 04/12}
\vspace{9.5cm}

\begin{abstract}
We study D2-branes on the K3-fibration $\PP^4_{(11222)}[8]$ using matrix factorizations at the Landau-Ginzburg point and analyze their moduli space and superpotentials in detail. We find that the open string moduli space consists of various intersecting branches of different dimensions. Families of D2-branes wrapping rational curves of degree one intersect with bound state branches.
The influence of non-toric complex structure deformations is investigated in the Landau-Ginzburg framework, where these deformations arise as bulk moduli from the twisted sectors.
\end{abstract}

\section{Introduction}

D-branes in Calabi-Yau compactifications are interesting for various reasons.
On the one hand, space-time filling D-branes provide a method to embed non-abelian gauge theories into string theories. The low energy physics is  determined by the geometry of the D-brane and the background compactification. Here, the superpotential as well as the moduli space of supersymmetric vacua are key properties of the four-dimensional theory. On the other hand, D-branes in Calabi-Yau compactifications are also very interesting from a purely mathematical point of view, they are the subject of open string mirror symmetry and provide the possibility to calculate disk instantons. Also in this context, it is the superpotential that plays a dominant role.

In this paper, we study D-branes in a particular Calabi-Yau compactification, namely the K3-fibration $\PP^4_{(11222)}[8]$. Our investigation starts at the Landau-Ginzburg point in K\"ahler moduli space.
We focus on B-type branes, which are described by matrix factorizations of the Landau-Ginzburg superpotential. The simplest branes in the class we consider correspond geometrically to D2-branes wrapping rational curves of degree one of the Calabi-Yau. As it turns out, the moduli space of such branes has several interesting features.

First of all, at the Fermat point of the Calabi-Yau, there exist families of D-branes wrapping rational curves, generated by open string moduli. The moduli spaces of the D2-branes we consider are given by Riemann surfaces.
For a matrix factorization describing such a D2-brane, one finds one unobstructed complex deformation parameter that determines the position along the surface, as expected from geometry. However, this provides just one single branch
of the full moduli space of the brane; generically, this branch can intersect with other branches. We show that on other branches the number of unobstructed open string deformations can be different, as is compatible with $\mathcal{N}=1$ supersymmetry in four dimensions. In the example at hand, D2-brane branches with a single deformation parameter can intersect with moduli space branches with two deformation parameters. 
As it turns out in our model, the brane on the two-moduli branch can be decomposed into two constituent branes with an open string tachyon turned on. The two moduli correspond to the motion of the two constituent branes that form the bound state. These remain unobstructed even in the presence of the tachyon. 
At the Landau-Ginzburg point, which is a $\ZZ_8$ orbifold point, the two constituent branes just differ in the representation label of the orbifold group. At large volume, one of the constituent branes can be interpreted as a D2, the other is then the image of that D2 under the Gepner monodromy.

Interestingly, one can already ``see'' from the structure on the one-modulus branch that there might be an intersection with a two-moduli branch. To be more precise, on the one-modulus branch there are, apart from the unobstructed deformation parameter generating the branch, two more marginal open string states, that are however obstructed. These then become truly marginal on the two-parameter branch, whereas the initial modulus on the one-modulus branch becomes obstructed.

\medskip
\medskip

We also investigate the behavior of the brane family under bulk deformations. Generically, as the complex structure is deformed, the open string moduli space gets lifted, and there are only finitely many branes surviving the bulk deformation. All other branes flow to these fixed points. The obstructions are encoded in the superpotential, which for B-type branes is independent of the K\"ahler moduli and hence can be calculated at an arbitrary point in the K\"ahler moduli space.
For toric complex structure deformations, the D2-brane superpotential can be computed geometrically as a chain integral.
It has  been shown in \cite{Baumgartl:2010ad} how to reproduce this result (to first order in the bulk, and all orders in the boundary couplings) from the Landau-Ginzburg point of view. We show that for the bound states arising in this paper the superpotential is simply the sum of the superpotentials on the constituent branes.

The additional feature arising in the model at hand is that there are non-toric complex structure deformations. From the Landau-Ginzburg point of view, these correspond to complex structure deformations coming from the twisted sector of the theory. It is expected from geometrical considerations \cite{Kachru:2000an} that certain brane families will exhibit superpotential terms coming from such complex structure deformations. To treat those, \cite{Kachru:2000an,Katz:1996ht,Candelas:1993dm} consider an alternative model, where these deformations appear as toric deformations.
In the Landau-Ginzburg model, calculating the superpotential amounts to computing world sheet correlation functions involving twisted sector bulk fields. This can be achieved by determining the bulk-boundary map and then calculating an open string correlator. It does in particular not require to find an alternative model; one can directly calculate the necessary correlation functions. Our results are in agreement with those of \cite{Kachru:2000an}.

This paper is organized as follows: We start out with a brief description of the most important geometric features of the main model considered in this paper in section~\ref{sec:model}. In section~\ref{sec:MF} we briefly review the necessary ingredients from matrix factorizations that are required for the analysis; for more extensive and detailed discussion, we refer to~\cite{Kapustin:2002bi,Brunner:2003dc,Lazaroiu:2003zi,Herbst:2008jq}. In~particular, we discuss bound states of D-branes in the matrix factorization framework and derive some general results that we make use of in the subsequent discussion.
In section~\ref{sec:branches} we consider the boundary moduli space of $\PP^4_{(11222)}[8]$. We study in detail both the one- and two-moduli branches and show how they intersect.
We also construct a moduli space of bound states in the quintic and give a comparison with $\PP^4_{(11222)}[8]$.
Section \ref{sec:Bulk-deformations} is devoted to the calculation of superpotential terms induced by bulk perturbations. We briefly summarize the results on ordinary non-twisted (toric) deformations in the model at hand. 
We also investigate twisted sector deformations and construct the bulk-boundary map for these twist fields. The superpotentials arising from these non-toric deformations are studied in more detail.

\section{The Model} \label{sec:model}

In this section we will give a brief geometric introduction to the branes considered in our model, and explain their main features in this language. Our  analysis later in the paper will however be non-geometric.

We start out by collecting some facts on the geometry of our model \cite{Candelas:1993dm}, which is a hypersurface $X$ given by the vanishing locus of the polynomial
\begin{equation}\label{eq:W11222}
W=x_1^8+x_2^8+x_3^4+x_4^4+x_5^4
\end{equation}
in $\PP^4_{(11222)}$. The projective space has singularities along $x_1=x_2=0$; this locus is intersected by the hypersurface along the curve
\begin{equation}\label{eq:singcurve}
x_1=x_2=0\ , \qquad x_3^4+x_4^4+x_5^4 = 0 \ .
\end{equation}
These singularities are resolved by blowing up each point in (\ref{eq:singcurve}) into a $\PP^1$, thereby replacing the curve of singularities by a divisor $E$. 
Other divisors can be described as vanishing loci of polynomials of definite degree; following the notation of \cite{Candelas:1993dm} we denote the divisor corresponding to polynomials of degree one by $L$ and the one corresponding to polynomials of degree $2$ by $H$. The divisors $H$ and $L$ generate $H_4(X,\ZZ)$ and are related to $E$ by $|H|=|2L+E|$. The intersection of two divisors on the hypersurface is generically a 2-cycle. 
Here we define the 2-cycles $h$ and $l$ as the intersections
\begin{equation}
4l = H\cdot E \quad , \quad 4h=H\cdot L \ .
\end{equation}
Using $|H|=|2L+E|$ we then also have
\begin{equation}
H^2= 4l + 8h \ .
\end{equation}
The intersection relations between the curves and divisors are
\begin{equation}
H\cdot h=1, \quad L \cdot l = 1, \quad H\cdot l =0, \quad L\cdot h =0 \ ,
\end{equation}
hence $h$ and $l$ are dual to $H$ and $L$, and they generate $H_2(X,\ZZ)$. We are interested in D2-branes, hence branes that carry any combination of the charges $l$ and $h$. Accordingly, the moduli space of all D2-branes falls into different disconnected components, classified by K-theory. In this paper we are interested in the parts of low D2-brane charges. 

\medskip
\medskip

\noindent
Our main focus is on D2-branes wrapping rational curves of degree 1, i.e.:
\begin{equation}\label{eq:rationalcurve}
\mathbb{P}^1 \hookrightarrow \mathbb{P}^4_{(11222)}  \qquad (u,v) \rightarrow (f_1(u,v),\ldots,f_5(u,v)),
\end{equation}
where $f_{1,2}$ and $f_{3,4,5}$ are homogeneous polynomials of degree one and two, respectively, and  $(u,v)$ parametrize the curve. Intersecting this set with the hypersurface specified by $W=0$ imposes a constraint on the polynomials $f_i$. As a  simplification, we consider the case where $2$ out of the $5$ coordinates $x_i$ are proportional to $u$, and $3$ are proportional to $v$, in formulas, we specialize to the following embeddings:
\begin{equation}\label{eq:ansatzcurve}
(x_{i_1}, \; x_{i_2})=(u^{k_{i_1}}, \; \eta u^{k_{i_2}}), \qquad (x_{i_3}, \; x_{i_4}, \; x_{i_5})=(a v^{k_{i_3}}, \; b v^{k_{i_4}}, \; c v^{k_{i_5}}),
\end{equation}
with $k_{i_j}=1$ or $2$. The hypersurface equation yields a constraint on $\eta$, as well as a condition on the $3$ complex parameters $(a,b,c)$. Taking into account projective equivalence, (\ref{eq:ansatzcurve}) yields a complex one-parameter family of embedded spheres. It is easy to see that there are altogether $\binom{5}{2} = 10$ such families falling into 3 types (corresponding to 3 possible ways to combine the weights of $(x_{i_1}, x_{i_2})$). In section \ref{sec:branches}, we will study each type separately and determine its intersections with other branches, which occur at special points in the respective moduli spaces. Branches can only intersect if this is compatible with K-theory, and the rational curves (\ref{eq:ansatzcurve}) fall into two different classes. 

Let us illustrate this with a concrete example, namely the curves given by
\begin{equation}\label{eq:example-branch}
(x_1,x_2,x_3,x_4,x_5) = (\eta u, u, av^2, bv^2, cv^2) \ ,
\end{equation}
where the parameters are subject to the constraint
\eqn{
a^4+b^4+c^4=0, \qquad \eta^8=-1 \ .
}
The moduli space is a Riemann surface of genus $2$. An intersection with another family of curves can be found, e.g., at the point
\eqn{
(x_1,x_2,x_3,x_4,x_5) = (\eta u, u, \eta' v^2, 0, v^2), \qquad (\eta')^4=-1 ,
}
which is also part of the family
\eqn{
(x_1,x_2,x_3,x_4,x_5) = (a'v, c'v, \eta' u^2, b' v^2, u^2) .
}
Both branches are single D2-brane branches of the type (\ref{eq:ansatzcurve}), where the moduli space on each branch is given by a Riemann surface. The two Riemann surfaces share one point. As we shall see explicitly in the Landau-Ginzburg discussion, the open string spectrum of the curves contain two marginal operators, of which only one is truly marginal at generic points. The other is obstructed and can be interpreted as the deformation along the other branch. At the intersection point, both operators are truly marginal, as at this point one is free to move in either direction. 

We can easily calculate the charge of the branes in this part of the moduli space. Consider the point where the two branches intersect, and note that the embedding of the sphere can be rewritten as the vanishing locus of the polynomial
\eqn{
J_1= x_1- \eta x_2, \quad J_2=x_4, \quad  J_3= x_3-\eta' x_5 \ .
}
The divisor corresponding to $J_1$ is in $|L|$, whereas the one corresponding to $J_2$ is in $|H|$. Thus, the intersection of $J_1$ and $J_2$ with the hypersurface yields $4$ curves of type $h$. The third equation $J_3=0$ singles out one of the curves, so that indeed the corresponding branes carry one unit of D2-brane charge.

However, there are also branes of different charge in the class (\ref{eq:ansatzcurve}). 
To see this, consider now the family
\eqn{
(x_1,x_2,x_3,x_4,x_5) = (\eta u, bv , a v^2, u^2 , cv^2) .
}
This branch can be described by the zero set of the linear functions
\eqn{
J_1=x_1^2-\eta^2 x_4, \quad J_2=ax_2^2 -b^2 x_3, \quad J_3= cx_3-ax_5 \ .
}
To calculate the charge, consider for example the point in moduli space where $b=0$, $c=1$, and $a=\eta'$, where $\eta'$ is a fourth root of $-1$. The intersection of $J_1, J_2$ and the hypersurface is then given by the intersection product $H\cdot H= 4l+8h$. This set decomposes into $4$ subsets, and $J_3=0$ describes one of them. Hence, the charge of the brane is $2h+l$ and the charge on this branch is not an elementary D2-brane unit. At the special point described above, the brane can be decomposed into a bound state of a single D2-brane of charge $h$ and another brane of charge $h+l$. This decomposition is however only possible at this very special point of the moduli space and does not persist at generic points.

Indeed, at the point in moduli space specified above, there is an intersection with a branch described by
\eqn{ \label{eq:bound_geo}
J_1'=x_3-\eta' x_5, \quad J_2'=(a')^2x_2^2-(c')^2 x_1^2, \quad J_3'= b'x_1^2-(a')^2 x_4 .
}
Note that $J_2'$ factorizes into two linear factors: $J_2'=(a'x_2+c'x_1)(a'x_2-c'x_1)$. Hence, this family is no longer of the type (\ref{eq:ansatzcurve}). The product form of $J_2'$ suggests that the branes in this family are bound states of two branes involving the two linear factors of $J_2'$. We shall confirm this expectation in the LG analysis by showing this decomposition on the level of the associated matrix factorizations.

There are two exactly marginal operators on the bound state branch (\ref{eq:bound_geo}), corresponding to the motion of the two constituents, so that the family can be extended to a two parameter family. In addition, there is one further marginal operator, that is however obstructed along the bound-state branch. On the intersecting single brane branch, the roles get exchanged, the previously obstructed deformation parameter becomes unobstructed, while there are two marginal but not truly marginal operators on that branch. This will be made quite explicit in the LG discussion, where the bound state branch is referred to as branch II*.

Summarizing, the open string moduli space of the families (\ref{eq:ansatzcurve}) falls into two disconnected components, distinguished by K-theory. The branches with non-elementary D2-brane charge intersect with bound state branches that are not of the type (\ref{eq:ansatzcurve}).

A further disconnected D2-brane family arises from D-branes wrapping the exceptional $\PP^1$s. This family has one parameter, corresponding to the position on the singular curve.

\medskip
\medskip

The above discussion is valid at the Fermat point. At generic points in the complex structure moduli space, there are no families of rational curves \cite{Candelas:1993dm}. For our model, there are altogether 86 possible complex structure deformations, 83 of which can be realized as monomials perturbing the defining equation. For these, it is easy to see that only certain specific members of a family survive the deformation. For concreteness, consider again the curves (\ref{eq:example-branch}). Perturbing the defining equation of the Calabi-Yau manifold 
\eqn{
x_1^8+x_2^8 + x_3^4+ x_4^4 + x_5^4 + s^{(8-d)}(x_1, x_2) s^{(d)}(x_3, x_4, x_5)=0,
}
with $s^{(d)}$ being a quasi-homogeneous polynomial of degree $d$, one sees that the ansatz~(\ref{eq:example-branch}) now only corresponds to curves on the hypersurface if
\eqn{
a^4 + b^ 4 +c^4= 0, \quad \text{and}  \quad s^{(d)}(a,b,c) =0 \ .
}
Thus, the 1-parameter family of curves collapses to a discrete set of curves. Physically, this is due to a superpotential generated when turning on the bulk deformation, and we will compute such superpotentials in section \ref{sec:Bulk-deformations}.

In addition, there are $3$ complex structure moduli that are non-toric and do not have a realization as monomials perturbing the defining equation. Rather, they are related to the blow-ups that are necessary to resolve the singularities of the weighted projective space. Turning on these moduli lifts the open string moduli space of the D2-branes wrapping the exceptional $\PP^1$s; again, a calculation of the corresponding superpotential can be found in section \ref{sec:Bulk-deformations}.

\section{Matrix Factorizations, Deformations and Bound States} 
\label{sec:MF}

At the stringy point of the K\"ahler moduli space, the internal part of a compactification on a Calabi-Yau hypersurface has a description in terms of an orbifold of a Landau-Ginzburg model, where the LG superpotential $W$ coincides with the defining polynomial of the hypersurface \cite{Witten:1993yc}.

In this section, we review some basics of the description of B-type D-branes in LG models in terms of matrix factorizations \cite{Kapustin:2002bi,Brunner:2003dc}. For later use in the paper, we in particular collect some facts from the cone construction as well as the necessary techniques to compute superpotentials. We use the same notation for superfields and geometric coordinates throughout the paper.

\subsection{Matrix Factorizations}

A matrix factorization of an LG superpotential $W$ is given by the data $(M, \sigma, Q)$, where $M$ denotes a free $\ZZ_2$-graded $\CC[x_i]$-module, $\sigma$ is the $\ZZ_2$-grading operator, and $Q$ is an odd operator with a matrix representation satisfying
\begin{equation}\label{def:MF}
Q^2 = W \cdot \mathbf{1}\ ,
\end{equation}
with $\mathbf{1}$ being the identity on $M$.
In the topologically B-twisted model $Q$ can be interpreted as the boundary part of the BRST operator; in particular the boundary spectrum consists of morphisms corresponding to fields that are $Q$-closed modulo $Q$-exact states. More precisely, starting with two matrix factorizations $(M_A, Q_A, \sigma_A)$ and $(M_B, Q_B, \sigma_B)$, the action of the BRST operator on a boundary field $\rho \in {\rm Hom}_{\CC[x_i]}(M_A,M_B)$ is given by
\eqn{
[Q_{\text{BRST}}, \rho ] = Q_B \rho-\sigma_B \rho \sigma_A Q_A \ .
}
We denote the cohomology of this operator by $H(Q_A,Q_B)$. The open string field $\rho$ is fermionic if it is an odd operator with respect to the $\ZZ_2$ grading, and bosonic if it is even.

In many examples, in particular those treated in this paper,  matrix factorizations can be expressed in terms of free boundary fermions $\pi^i$ and $\bar\pi^i$.
These fermions satisfy the standard Clifford algebra relations
\eqn{
	\{ \pi^i, \bar\pi^j \} = \delta^{ij}, \quad \{ \pi^i, \pi^j \} = 0, \quad \{ \bar\pi^i, \bar\pi^j \} = 0 \ .
}
In this language $M$ is $\CC[x_i]^{\oplus d}$, where $d=2^n$ is the dimension of the representation of the Clifford algebra with $n$ pairs of fermions. We can think of $M$ as a Fock representation obtained by applying creation operators $\bar\pi^i$ to a Fock vacuum $|0 \rangle $ that is annihilated by the $\pi^i$. The grading is  the natural  one on bosonic and fermionic subspaces. Given then a decomposition of $W(x_i)$ into a sum of products of polynomials $J_i$ and $E_i$
\begin{equation}
 W = \sum_{i=1}^n E_i J_i \ ,
\end{equation}
we can construct the boundary BRST-operator by taking
\eqn{
\label{defQ}
Q= \sum_{i=1}^n (J_i \pi^i + E_i \bar\pi^i),
}
which clearly fulfils (\ref{def:MF}).

The stringy points of Calabi-Yau compactifications given by hypersurfaces in weighted projective space are described by orbifolds of Landau-Ginzburg theories. The relevant orbifold group is $\ZZ_H$, where $H$ is the degree of the quasi-homogeneous polynomial $W$. In the bulk, the orbifold group acts by phase multiplication on the chiral superfields
\begin{equation}
x_i \rightarrow \omega_i x_i , \qquad \omega_i=e^{i \pi q_i}
\end{equation}
where $q_i=2w_i/H$ is the $U(1)$ charge of the chiral superfield, and $w_i$ denotes the weight of the coordinate in the underlying projective space.
As usual, orbifolding means that one has to specify a representation of the $\ZZ_H$ orbifold group on the Chan-Paton spaces.  In the context of matrix factorizations, this means that one considers BRST operators $Q$ with an equivariance condition:
\begin{equation}
\gamma Q(\omega_i x_i) \gamma^{-1} = Q(x_i) \ ,
\end{equation} 
where $\gamma$ is a representation matrix of the orbifold group. 
Open string operators $\Phi \in H(Q_A,Q_B)$ are projected on invariant morphisms by requiring that
\begin{equation}
\gamma_B \Phi(\omega_i x_i) \gamma_A^{-1} = \Phi \ .
\end{equation}
The $\ZZ_H$ representations are specified by a grade $\varphi \in \ZZ_H$, and the R-charge of a boundary field surviving the projection is given by 
\eqn{ \label{eq:orbgrading}
q_\Phi = \frac{2}{H}(\varphi_B-\varphi_A)+{\rm deg}(\Phi),
} 
where ${\rm deg}(\Phi)$ is the $\ZZ_2$ grade of $\Phi$.

\subsection{Effective superpotentials}

The effective space-time superpotential can be interpreted as a generating functional for the topological world sheet correlation functions (summed over operator orderings). One approach to compute it is therefore to directly calculate the relevant world sheet correlation functions and to determine the generating function. In the context of matrix factorizations, a second approach is to add perturbations to the equation $Q^2=W$, either by modifying $Q$ (boundary perturbations) or by perturbing $W$ (bulk perturbations). In this paper we will make use of both of these approaches geared to our specific situation; see \cite{Carqueville:2011aa} for a more complete and general discussion, as well as a comparison of different approaches to compute superpotentials perturbatively. 

Calculations of disk world sheet correlation functions with up to three boundary insertions can be performed using the Kapustin-Li formula \cite{Kapustin:2003ga}.
It is expressed as a residue integral around the critical points of the superpotential. In the matrix representation\footnote{In the fermion representation, the supertrace is replaced by a Grassmann integral over the boundary fermions.} it is given by
\begin{equation}
\langle \psi_1 \psi_2 \psi_3 \rangle = {\rm Res} \ \frac{{\rm STr}\ [(\partial Q)^{\wedge N} \psi_1 \psi_2 \psi_3]}{\prod_{i=1}^N \partial_i W} \ ,
\end{equation}
where $\psi_i$ denote open string operators in the respective cohomologies of the BRST operators of the branes involved, the supertrace is ${\rm STr}={\rm Tr}\sigma$, and $N$ is the number of variables, which is $5$ for the models considered in this paper.
The open string operators need to be compatible with the orbifold projection. The correlation function is then simply a projection of that of the unorbifolded theory, so that the above expression for the correlator holds without further modification.

Note that the Kapustin-Li correlator can also be applied to the case of one bulk and one boundary insertion. This is possible whenever the bulk-boundary map of the bulk field is known. 

In our case, there are two classes of bulk deformations in the topological sector. Bulk perturbations from the untwisted sector are realized in terms of monomial perturbations of the superpotential; in this case the bulk-boundary map takes the monomial simply to an identity matrix of appropriate dimension times itself. If the perturbation is from the twisted sector, selection rules can help to determine the image of the bulk-boundary map. The correlation function between a bulk field $\Phi$ and a boundary field $\psi$ is then
\begin{equation}\label{eq:Kapustin-li}
\langle \Phi \psi \rangle = {\rm Res} \ \Phi \frac{{\rm STr} \ [(\partial Q)^{\wedge N} \psi]}{\prod_{i=1}^N \partial_i W} \ .
\end{equation}

A priori, this correlator gives a contribution to the superpotential that is only first order in bulk and boundary couplings. However, for BRST operators forming continuous families, it is possible to calculate the correlation functions at any point of the family, thereby obtaining a result that is exact in the boundary couplings and first order in the bulk couplings.

To this end, let us assume that the BRST operator depends on  parameters $b_n$ and hence physically describes families of D-branes.
The open string states that generate such families can be obtained as derivatives of the BRST operator
\begin{equation}
\psi_{b_n} = \partial_{b_n} Q(b_n) \ .
\end{equation}
Note that since $Q^2(b_n)=W\cdot 1$, and $W$ does not depend on $b_n$ by definition, this operator is always $Q$-closed. Furthermore, it always has $U(1)$ charge $1$, since $Q$ has charge $1$ and hence corresponds to a marginal operator. 

Consider now turning on a bulk deformation. If the perturbation is in the untwisted sector, this is realized by perturbing $W$ by adding a polynomial. In general, $Q(b_n)$ will no longer provide a matrix factorization for the perturbed theory. From a geometric point of view, we have changed the complex structure, so that curves that initially were holomorphic are no longer holomorphic after the  perturbation. From a physical point of view, this lifting of the moduli space can be attributed to a superpotential. This superpotential can now be computed to first order in the bulk coupling using (\ref{eq:Kapustin-li}) at any value for $b_n$. Since the complete $b_n$-dependence of the correlator is known, the correlation function is known to any order in the boundary couplings. A further integration with respect to the boundary parameters $b_n$ yields the superpotential.

\medskip
\medskip

As mentioned at the beginning of this section, a second approach to study the structure of the open string moduli space is to consider perturbations of $Q$.
Following \cite{Hori:2004ja} (see \cite{Ashok:2004xq,Herbst:2004jp,Jockers:2007ng,Aspinwall:2007cs,Knapp:2008tv,Knapp:2009vu,Carqueville:2011px} for related papers making use of perturbation theory in the context of matrix factorizations), we make the ansatz
\begin{equation}\label{eq:Qperturb}
Q'=Q+\sum_{n=1}^\infty \lambda^n Q_n \;,
\end{equation}
where the first order term $Q_1$ is a fermion in the cohomology of $Q$, as can be seen by demanding that $Q^2=W \mathbf{1}$ at first order. Requiring $(Q')^2=W \mathbf{1}$, at arbitrary order $\lambda^n$ gives the equations
\begin{equation}
\{Q,Q_n\}=-\sum_{k=1}^{n-1}Q_k Q_{n-k} \;.
\end{equation}
This system of equations can be solved iteratively; solving the system at order $n-1$ (choosing $Q_1, \dots, Q_{n-1}$) the above equation gives a condition on $Q_n$. Note that it does not necessarily have a solution,  the perturbation by $Q_1$ can have an obstruction at a certain finite order. In the case that there are no obstructions, the procedure yields a new family of  matrix factorizations, parametrized by $\lambda$, and $Q_1$ generates a branch in the moduli space. 

In this paper, we will only use the above perturbative procedure to determine whether a boundary perturbation is obstructed or not; see \cite{Carqueville:2011aa} for an algorithm that keeps the information of the obstruction at any order, thereby providing a method to compute effective superpotentials.

\subsection{Bound States}
\label{MF:boundstates}

Let us finally turn to the description of bound states of D-branes in the matrix factorization framework. 
The starting point are two matrix factorizations $(M_A,\sigma_A,Q_A)$ and $(M_B,\sigma_B,Q_B)$ together with an odd boundary changing operator $T\in H(Q_A,Q_B$ that we interpret as a (topological) tachyon. Note that if the two branes carry different $\ZZ_H$ representation labels, such that the open string operator $T$ has a charge smaller than one, then $T$ is indeed a tachyon in the physical theory.

In the matrix representation the bound state of the two matrix factorizations is given by $(M_A \oplus M_B, \sigma_A \oplus \sigma_B, Q)$, where
\begin{equation}\label{eq:bound_state}
Q=\begin{pmatrix}
Q_A & 0 \\
T & Q_B
\end{pmatrix},
\end{equation}
with $T$ being an odd operator in $H(Q_A,Q_B)$.
Mathematically this is referred to as the cone construction over B-brane categories forming a triangulated category.

Let us now collect a couple of results that we will make use of in our later discussion. First, consider the topological disk correlators for a $Q$ of the form (\ref{eq:bound_state}). It is easy to see that such correlators will be independent of the tachyon $T$, provided that the boundary insertions are also represented by lower triangular matrices.\footnote{This will be the case for almost all boundary fields we encounter in our examples, since the fields will be usually represented by linear combinations of $f_i \partial_{\beta_i} Q$, where $\beta_i$ are some boundary moduli, and $f_i$ are rational functions in $x_1,\ldots,x_N$. Since $Q$ is lower triangular, so will be the boundary fields.} In fact, if the boundary operator $\Psi$ is of the form
\eqn{
\Psi=\begin{pmatrix}
\Psi_A & 0 \\
X & \Psi_B
\end{pmatrix},
}
we find that
\eqn{
 \text{STr}[(\partial Q)^{\wedge N} \, \Psi]
 &=\text{Tr}[(\sigma_A \oplus \sigma_B)(\partial Q)^{\wedge N} \, \Psi]\\
 &=\text{STr}[(\partial Q_A)^{\wedge N} \, \Psi_A]+\text{STr}[(\partial Q_B)^{\wedge N} \,\Psi_B]\ .
}
For an arbitrary bulk insertion $\Phi$, we thus have
\begin{equation}
\label{corrsum}
\langle \Phi\Psi \rangle_{Q}=\langle \Phi \Psi_A \rangle_{Q_A}+\langle \Phi \Psi_B \rangle_{Q_B} \ ,
\end{equation}
which immediately implies that the effective superpotential does not contain a term coming from the tachyon $T$. 

Another simple observation is the following. 
Consider two matrix factorizations $Q_A$, $Q_B$ that can be decomposed as (graded) tensor products
and share a common factor,~i.e.
\eqn{\label{eq:tensordecomp}
Q_A= Q_1 \otimes Q_{2A}, \qquad Q_B = Q_1 \otimes Q_{2B} \ .
}
The cohomology decomposes accordingly as
\eqn{
H(Q_A, Q_B) = H (Q_1, Q_1) \otimes H(Q_{2A}, Q_{2B}) \ .
}
Choosing then a tachyon of the form $1 \otimes T \in H(Q_A, Q_B)$, it is not difficult to see that we have the following equivalence of matrix factorizations
\begin{equation}\label{eq:boundtensor}
Q=\begin{pmatrix}
Q_1 \otimes Q_{2A} & 0 \\
1 \otimes T & Q_1 \otimes Q_{2B}
\end{pmatrix} \ \sim \ Q_1 \otimes
\begin{pmatrix}
 Q_{2A} & 0 \\
T & Q_{2B}
\end{pmatrix} .
\end{equation}
In other words, this way we build the bound state in the second factor only.

In our constructions, we will consider bound states of the form (\ref{eq:boundtensor}), where both $Q_{2A}$ and $Q_{2B}$ will depend on one complex boundary modulus $\beta_A$ and $\beta_B$, respectively. Taking $T=0$, one can then trivially obtain a two-dimensional family of bound states, which are simply a direct sum of the constituents. However, for our examples, we will be able to construct such a 2-dimensional family even for a non-trivial $T$. More specifically, in our case the matrix factorizations will have the following form in the fermionic representation
\begin{equation}\label{eq:fermrepr}
 Q_{1}=J_{1} \pi^1+E_{1} \bar{\pi}^1, \qquad Q_{2A}=J_{2A} \pi^2 + E_{2A} \bar{\pi}^2+J_{3A} \pi^3 + E_{3A} \bar{\pi}^3,
\end{equation}
with an analogous expression for $Q_B$. We then take the following ansatz for the tachyon
\eqn{
T= \pi^2 + T_1 \pi^3 + T_2 \bar\pi^2 + T_3 \bar\pi^3,
}
i.e., we take only terms linear in the boundary fermions $\pi$, with the coefficient of one of the terms being constant. It is easy to see that the condition for $T$ to be closed, $Q_{2B} T + T Q_{2A}=0$, implies
\eqn{
\label{Tachyon}
	T_1=\frac{J_{3A}-J_{3B}}{J_{2A}-J_{2B}}, \qquad T_2=\frac{E_{2A}-E_{2B}}{J_{2A}-J_{2B}}, \qquad T_3=\frac{E_{3A}-E_{3B}}{J_{2A}-J_{2B}}\ .
}
This is just a formal solution, and in general, one has to make sure that these expressions are polynomial and that $T$ is not trivial in $H(Q_{2A}, Q_{2B})$. As we will see shortly, $T$ will have both of these properties in our examples.\footnote{It will actually turn out that $T$ is the only element in $H(Q_{2A}, Q_{2B})$.} Since the tachyon can be constructed for any value of the moduli $\beta_A$, $\beta_B$, the resulting family of bound states (\ref{eq:boundtensor}) will indeed be two-dimensional.

\section{The Open String Moduli Web} 

\label{sec:branches}

In this section, we discuss the various branches of the D2 moduli space of $\mathbb{P}^4_{(11222)}[8]$ in detail. Our analysis will be based on the study of the associated matrix factorizations at the Landau-Ginzburg point. These matrix factorizations can be found by noting that the curves~(\ref{eq:ansatzcurve}) can be equivalently described as vanishing loci of parameter dependent polynomials $J_i(x_j)$. One can then construct polynomials $E_i(x_j)$ such that $W$ can be decomposed as $W=\sum_{i=1}^3 J_i(x_j) E_i (x_j)$ and apply the formalism of section \ref{sec:MF} to convert this into a matrix factorization. 

Note that at this level this is just a formal prescription to associate matrix factorizations to geometric objects. To connect D-branes in the Landau-Ginzburg description to geometric objects, one has to apply the transport to large volume worked out in \cite{Herbst:2008jq}. Of course, the transport is in general path dependent, as one can for example circle singular loci in the K\"ahler moduli space. Furthermore, the Landau-Ginzburg branes come with a grade, and D-branes of different grade (but equivalent matrix factorizations) correspond to different Landau-Ginzburg monodromies of the same brane at large volume. The transport of branes described by matrix factorizations of the same type as the ones appearing in this paper has been discussed in section~10 of \cite{Herbst:2008jq} (for the case of the quintic) and for our model in \cite{Aspinwall:2006ib}.  Indeed, one of the branes obtained from these matrix factorizations is a D2-brane specified by the vanishing locus of linear maps $J_i$. For the purpose of this paper, we have verified on the level of charges that the matrix factorizations obtained by the above prescription indeed fall into the expected large volume K-theory classes, see section~\ref{sec:moduliweb} for further comments.
\medskip

After studying each moduli branch separately we examine the global structure of the moduli space, in particular, the intersection points of the individual branches in the moduli space. Interestingly, investigation of the moduli space of D-branes wrapping~(\ref{eq:ansatzcurve}) reveals the existence of one- and two-dimensional families of D-branes continuously connected to this moduli space. As we shall see shortly, these 1d- and 2d-branches are related to bound states of the D-branes wrapping~(\ref{eq:ansatzcurve}). The branches can be separated into various types, which are summarized in table~\ref{tab:modspaces1}. We find three different 1d-branches together with starred branches, that are associated to bound states of 1d-branches. One of the two starred branches is a 2d-branch.

The bound states branches that we find in $\mathbb{P}^4_{(11222)}[8]$ can be constructed in a very similar way also for the quintic. We find a similar heterogeneous web of mixed dimension, while transitions from single branes to bound states are absent.

\begin{table}
\begin{center} \renewcommand{\arraystretch}{1.15}
\begin{tabular}{|c|c|c|}
\hline
\multirow{2}{*}{Type} & \multicolumn{2}{|c|}{Moduli space $\cal M \times \cal N$}\\ \cline{2-3}
& $\cal M$ & $\cal N$\\
\hline
I& $\mathcal{M}_I= \{a^4+b^8+c^4=0 \subset \mathbb{P}^2_{(212)} \}$ 
	& ${\cal N}_I=\{\eta^8=-1\}/\ZZ_2$ \\ \hline
II& $\mathcal{M}_{II}= \{a^8+b^4+c^8=0 \subset \mathbb{P}^2_{(121)}\}$
	& ${\cal N}_{II}=\{\eta^4=-1\}$ \\ \hline
III& $\mathcal{M}_{III}= \{a^4+b^4+c^4=0 \subset \mathbb{P}^2_{(111)}\}$
	& ${\cal N}_{III}=\{\eta^8=-1\}$ \\ \hline
II* & $\mathcal{M}_{II}\times \mathcal{M}_{II}$ 
	& ${\cal N}_{II}$ \\ \hline
III* & $\mathcal{M}_{III}$ 
	& $\{(\eta,\eta'), \eta^8=\eta'^8=-1, \eta \neq \eta'\}/S_2$ \\ \hline
\end{tabular}\caption{Types of branches with the corresponding moduli spaces. The branches II* and III* describe D-branes that are bound states of D-branes in the 1d-branches II and III, respectively. Branch II* is a 2d-branch while III* is a bound state between a 1d-branch and a copy of that branch at a different value of the discrete modulus $\eta$, therefore it is also 1-dimensional.\label{tab:modspaces1}}
\end{center}
\end{table}

\subsection{Moduli Branches of the Model \texorpdfstring{$\mathbb{P}^4_{(11222)}[8]$}{P11222[8]}}
\label{sec:Branches_P11222}

\subsubsection{Branch I}

Let us start our discussion by studying the following family of rational curves:\footnote{See \cite{Alim:2010za} for a discussion of curves in this family from a geometric persepctive.}
\begin{equation}\label{eq:BranchI_curves}
(x_1,x_2,x_3,x_4,x_5)=(\eta u, b v, a v^2,  u^2, c v^2), \qquad (u,v)\in \mathbb{P}^1 \ ,
\end{equation}
where in order to satisfy $W=0$, we have to require
\begin{equation}\label{eq:BranchI_cond}
 a^4+b^8+c^4 =0, \qquad \qquad \eta^8=-1 \ .
\end{equation}
Note that reparametrizations of the curves allow us to identify $\eta \sim -\eta$, and take $(a,b,c) \in \mathbb{P}^2_{(212)}$. Equation (\ref{eq:BranchI_cond}) thus defines four copies of a Riemann surface corresponding to the moduli space of (\ref{eq:BranchI_curves}). The matrix factorizations associated to this family of D2-branes are given by $Q(a,b,c)=\sum_{i=1}^3 (J_i \pi^i + E_i \bar{\pi}^i)$ with\footnote{It is sometimes useful to write $E_i$ in the following form:\\
$E_1=\prod_{\eta' \in I_1}(x_1^2-\eta'^2 x_4),\ \quad 
E_2=\frac{1}{a^{4}}\prod_{\eta' \in I_2}(a x_2^2-(\eta' b)^2 x_3),\ \quad 
E_3=-\frac{1}{a^{4}}\prod_{\eta' \in I_3}(c x_3-\eta' a x_5)\ $
with\\
$I_1=\{ \eta'^8=-1, \eta'^2 \neq \eta^2 \}/\ZZ_2,\ \quad 
I_2=\{\eta'^8=1, \eta'^2 \neq 1 \}/\ZZ_2, \ \quad 
I_3=\{\eta'^4=1, \eta' \neq 1 \}.$
}

\begin{align}\label{eq:mfbranchI}
\begin{array}{ll}
J_1= x_1^2 -\eta^2 x_4 & \qquad \qquad \qquad E_1 = \sum_{n=0}^3(x_1^2)^n(\eta^2 x_4)^{3-n}\\
J_2 = a x_2^2 - b^2 x_3 & \qquad \qquad \qquad E_2 = \sum_{n=0}^{3}\frac{(a x_2^2)^n(b^2 x_3)^{3-n}}{a^4} \\
J_3 = c x_3 - a x_5 & \qquad \qquad \qquad E_3 = \sum_{n=0}^{3}\frac{(c x_3)^n(a x_5)^{3-n}}{-a^4}\; .
\end{array}
\end{align}
One can easily check that the matrix factorization condition $Q^2=W \mathbf{1}$ is satisfied provided that (\ref{eq:BranchI_cond}) holds. Note that here we assume $a \neq 0$, but one can proceed similarly also in the other two patches $b \neq 0$, and $c \neq 0$.  At the common points, the corresponding factorizations are gauge equivalent. In what follows, we will continue to work in the patch $a \neq 0$.

In order to understand the structure of the open string moduli space, one has to study boundary perturbations generated by marginal operators in the boundary preserving spectrum, i.e., R-charge 1 elements in the cohomology of $Q$. It turns out that at a generic point of the branch there are 3 fermions in the marginal spectrum, with the following representatives:\footnote{Note that $\partial_b Q$ is well-defined only for $c \neq 0$. The representatives in the other patches can be constructed in a similar way.}
\begin{equation}
\chi=\frac{1}{2b}\partial_b Q, \qquad \tilde \psi_1= \frac{x_1^2}{x_3} \chi, \qquad \tilde \psi_2= \frac{a x_1 x_2}{x_3} \chi \; .
\end{equation}
The field $\chi$ obviously corresponds to the exactly marginal field generating translations along this branch. From~(\ref{eq:mfbranchI}) one can also see that $\chi \propto x_3$, which enables us to write $\tilde \psi_1$ and $\tilde \psi_2$ as above in a well-defined way. In order to see if $\tilde \psi_1$ and $\tilde \psi_2$ are obstructed, we compute the relevant correlators. Using the Kapustin-Li formula one finds that the only non-vanishing three-point function is
\begin{equation} \label{eq:3p1}
\langle \tilde{\psi}_1 \tilde{\psi}_1  \tilde{\psi}_2 \rangle= \frac{3 a^3 b^4 \eta^6}{8 c^7} \;.
\end{equation}
This (and a similar calculation in the other patches) implies that $\tilde \psi_1$ is obstructed at first order except for the points where $a$, $b$, or $c$ are zero. As we will see later, these are precisely the permutation points corresponding to intersections with other branches. At these points $\tilde \psi_1$ becomes a generator of the corresponding branch. To see if $\tilde \psi_2$ can generate a finite boundary perturbation, we treat the associated deformation of $Q$ perturbatively, as in (\ref{eq:Qperturb}).
For the present case, we find an obstruction at order 4, except for the point $b=0$ where $\tilde \psi_2$ is unobstructed.

We have thus found that, apart from the field $\chi$ generating translations along this branch, there are two additional fields, which are unobstructed only at certain special permutation points. In particular, at the points $a=0$ and $c=0$ only $\tilde \psi_1$ is unobstructed, and one can easily see that these are the points corresponding to intersections with other branches of type~I. On the other hand, at $b=0$ both $\tilde \psi_1$ and $\tilde \psi_2$ are unobstructed, and thus we might expect to find an intersection with a two-dimensional family of D-branes at this point. As we will see this is indeed the case: branch I intersects with II*, which is a 2d-branch of bound states of D-branes in branch II.

\subsubsection{Branch II}

This branch describes D-branes 
wrapping the following family of curves:
\begin{align}
(x_1,x_2,x_3,x_4,x_5)&=(a v, c v, \eta u^2, b v^2, u^2), \qquad (u,v)\in \mathbb{P}^1 \;,\\
a^8&+b^4+c^8 =0, \qquad \qquad \eta^4=-1, \label{eq:mod_spaceII}
\end{align}
with the associated matrix factorization being given by $Q=\sum_{i=1}^3 (J_i \pi^i + E_i \bar{\pi}^i)$ with
\begin{align}\label{eq:mfbranchII}
\begin{array}{ll}
J_1 = x_3 -\eta x_5 &  \qquad \qquad \qquad E_1 =\sum_{n=0}^{3} (x_3)^n (\eta x_5)^{3-n}\\
J_2 = a x_2 - c x_1 & \qquad \qquad \qquad E_2 = \sum_{n=0}^{7}\frac{(a x_2)^n (c x_1)^{7-n}}{a^8}  \\
J_3 = b x_1^2 - a^2 x_4  & \qquad \qquad \qquad E_3 = \sum_{n=0}^{3}\frac{(b x_1^2)^n (a^2 x_4)^{3-n}}{-a^8}.
\end{array}
\end{align}
The marginal fermionic spectrum consists only of $\theta=\partial_b Q$, except for the point $b=0$, where we have an additional fermion $\rho$ corresponding to the generator of branch III. This fermion exists only at $b=0$ and cannot be continued to other points on the branch.

\subsubsection{Branch II*}\label{sec:BranchII*}

We will now study a two-dimensional family consisting of bound states of D-branes in branch II, which will intersect with branch I at a finite number of points. Consider thus two matrix factorizations $Q_A$ and $Q_B$ of the form (\ref{eq:mfbranchII}) with boundary moduli $(a,b,c)$ and $(a',b',c')$, respectively. We construct a bound state of $Q_A$ and $Q_B$:
\begin{equation}\label{eq:Q_bound_eta}
Q=\begin{pmatrix}
Q_A & 0 \\
T & Q_B
\end{pmatrix},
\end{equation}
with $T$ being a boundary changing fermion in the cohomology $H(Q_A,Q_B)$. As we shall see shortly, for a certain choice of the fermion, we will this way obtain a two-dimensional family intersecting with branch I.

To specify the form of $T$, let us first inspect the structure of $H(Q_A,Q_B)$. We shall make use of the fact that $Q_{A}$ and $Q_B$ can be decomposed as
\begin{equation}\label{eq:Q_tensor_decomp}
 Q_A=Q_{1A}(x_3,x_5) \otimes Q_{2A}(x_1,x_2,x_4),
\end{equation}
with
\begin{equation}
 Q_{1A}=J_{1A} \pi^1+E_{1A} \bar{\pi}^1, \qquad Q_{2A}=J_{2A} \pi^2 + E_{2A} \bar{\pi}^2+J_{3A} \pi^3 + E_{3A} \bar{\pi}^3,
\end{equation}
and analogously for $Q_B$. Note that we take $Q_{1A}=Q_{1B}\equiv Q_1$. In view of (\ref{eq:Q_tensor_decomp}), the cohomology $H(Q_A,Q_B)$ admits the following decomposition: 
\begin{equation}
 H(Q_A,Q_B)=H(Q_1,Q_1)\otimes H(Q_{2A},Q_{2B}) .
\end{equation}
The $H(Q_1,Q_1)$ part is purely bosonic~\cite{Brunner:2005fv}, with a basis given by $\{1,x_3,x_3^2\}$. The fermionic part of $T$ must thus come from $H(Q_{2A},Q_{2B})$. For generic values of the moduli, there is precisely one fermion $T_f$ in $H(Q_{2A},Q_{2B})$, and it has $U(1)$ charge $3/4$.\footnote{There are additional fermions at the points where $a=a'$, $c=c'$.} The most general fermionic element in $H(Q_A,Q_B)$ has thus the form:
\begin{equation}
T=T_b \otimes T_f,
\end{equation}
with $T_b=\{1, x_3, x_3^2\}$. For our construction, we will make the choice $T = 1 \otimes T_f$. 

It should be clear that since $T$ can be found in $H(Q_{A},Q_{B})$ for any value of the boundary moduli $(a,b,c),(a',b',c')$ (provided they satisfy (\ref{eq:mod_spaceII})), we obtain a family of bound states (\ref{eq:Q_bound_eta}) that is (complex) two-dimensional. We now wish to show that this family intersects with branch I. Here and in the following we will work in the patch where $a \neq 0$ and $a' \neq 0$; by rescaling we can then set $a=a'$. In this patch a representative for $T_f$ can be constructed as follows (here we assume in addition that $c\neq c'$ or $b=b'$):
\begin{equation}\label{eq:tachyonII*_repr}
 T_f=\pi^2 +  T_1 \; \pi^3 + T_2 \; \bar{\pi}^2 +  T_3 \; \bar{\pi}^3 \;,
\end{equation}
with
\begin{equation}
 T_1=\frac{J_{3A}-J_{3B}}{J_{2A}-J_{2B}}, \qquad T_2=\frac{E_{2A}-E_{2B}}{J_{2A}-J_{2B}}, \qquad T_3=\frac{E_{3A}-E_{3B}}{J_{2A}-J_{2B}} \;.
\end{equation}
Bringing (\ref{eq:Q_bound_eta}) to the gauge equivalent form (see Appendix \ref{sec:App_matrices} for details):
\begin{align}
Q \sim \hat Q =& J_1 \pi^1 + E_1 \bar\pi^1 + J_{2A} J_{2B} \pi^2 - (T_2+ T_1 T_3) \bar \pi^2+ \nonumber\\
&+ (J_{3A} - T_1 J_{2A})\pi^3 + (E_{3A} - T_3 J_{2A}) \bar \pi^3 \;, \label{eq:Q_bound_gauge}
\end{align}
it is then not difficult to check that for $b=b', c=c'=0$, the bound state describes a D-brane in branch I.\footnote{In the calculation one can use that $\lim_{\substack{b' \rightarrow b\\ c' \rightarrow c}}\frac{b-b'}{c-c'}=
-2\frac{c^7}{b^3}$.}

Two remarks are in order. First, since we are dealing with an LG orbifold, and considering the boundary changing spectrum between $Q_A$ and $Q_B$, we have to specify the $\mathbb{Z}_8$ representation labels of these factorizations.
In particular, we should choose the representation labels in such a way that $T$ will remain in the spectrum after the orbifold projection. If we denote the labels by $\varphi_A$ and $\varphi_B$, we find from (\ref{eq:orbgrading}) that this gives the condition $\varphi_A-\varphi_B=1 \mod 8$. The second thing to notice is that (\ref{eq:tachyonII*_repr}) is well-defined only for $c\neq c'$ or $b=b'$. For $b \neq b'$ or $c=c'$, we can instead take $\frac{c-c'}{b-b'}T_f$. It can be easily seen that on the overlap, the resulting bound states will be gauge equivalent.

Having constructed the two-dimensional family of bound states (\ref{eq:Q_bound_eta}) and established the connection with branch I, we can now again focus on the possible boundary perturbations. The marginal boundary preserving spectrum on this branch consists of two fermions for a generic point $(a,b,c,a',b',c')$. These are precisely the fermions generating translations along the branch and one can easily construct their representatives by taking $\psi_1=\partial_c Q$, and $\psi_2=\partial_{c'} Q$. One can also find three additional marginal fermions in the spectrum, which we denote by $\chi$, $\omega$, and $\tau$. These exist however only on a certain subset of the two-dimensional moduli space. The field $\chi$ lives at the points $(a',b',c')=(a,b,-c)$, and it can be interpreted as a continuous extension of the generator of branch I. This relation can be seen by taking the gauge equivalent form (\ref{eq:Q_bound_gauge}), in terms of which $\chi$ can be written as $\chi = \frac{x_3}{2 c x_1^2}(\partial_c \hat Q - \partial_{c'} \hat Q)$, and comparing the expressions at the intersection point. It is important to note that the relation between the moduli holds also the other way round,~i.e., the fermions $\tilde \psi_1$, $\tilde \psi_2$ living on branch I are continuous extensions of the generators of branch II*. To be more precise, these fermions are related to the following linear combinations of~$\partial_c \hat Q$~and~$\partial_{c'} \hat Q$:
\begin{eqnarray}
 \tilde \psi_1 &=& \frac{\partial_c \hat Q - \partial_{c'} \hat Q}{c-c'},\\
 \tilde \psi_2 &=& \frac{\partial_c \hat Q + \partial_{c'} \hat Q}{2}.
\end{eqnarray}
The other marginal fermion $\omega$ can be found at the points $(a,b,c,a',b',c')=(1,0,\eta',1,0,\eta'')$, with $\eta' \neq \eta''$ being eighth roots of $-1$. These points are precisely the intersection points with branch III*, and $\omega$ is just the modulus of this branch. As we will see shortly, there are two additional marginal fields on branch III*, $\psi_1'$ and $\psi_2'$, which can be again seen as continuous extensions of generators of branch II*, namely
\begin{eqnarray}
 \psi_1' &=& -(\partial_b \hat Q + \partial_{b'} \hat Q),\\
 \psi_2' &=& -(c\; \partial_b \hat Q + c'\; \partial_{b'} \hat Q).
\end{eqnarray}
Let us note that $\chi$ and $\omega$ are actually related as they correspond to the same operator at the points $(1,0,\eta',1,0,-\eta')$. The last fermion that can be found in the marginal spectrum on this branch, $\tau$, lives at the points $b\neq b',c=c'$. Its representative can be constructed as follows:
\begin{equation}
 \tau=
x_3 \begin{pmatrix}
-\frac{1}{x_1}T & (b'-b) T\\
 0 & \frac{1}{x_1}T
\end{pmatrix},
\end{equation}
with $T$ being the tachyonic operator in (\ref{eq:Q_bound_eta}). As we will see shortly, unlike $\chi$ and $\omega$, this fermion cannot be interpreted as originating from a modulus of another branch.

We end this section by computing the possible three-point functions between the marginal fields. First, we note that $\psi_1$ and $\psi_2$ are (obviously) unobstructed on the entire family, hence any three-point function containing only these fields will vanish. Furthermore, since $\omega$ is unobstructed at the points where it exists, the only possibly non-vanishing correlator could be $\langle \psi_1 \psi_2 \omega \rangle$, but this can be found to be zero as well. For correlators involving $\chi$ we find that the only non-vanishing term is
\begin{equation}\label{eq:3p2}
\langle \chi \chi  \tilde{\psi}_2 \rangle= \frac{3 a^5 c^4 \eta^3}{8 b^7} \;.
\end{equation}
This shows, in particular, that $\chi$ is obstructed away from the intersection points with branch I and branch III*. Note as well that from (\ref{eq:3p1}),(\ref{eq:3p2}), and the vanishing of all other three-point functions, one can infer that the lowest order term in the effective boundary superpotential $\mathcal{W}(\tilde{\psi}_1,\tilde{\psi}_2,\chi)$ is
\begin{equation}
\mathcal{W}_0=k_1 \chi^4 \tilde{\psi}_2 \tilde{\psi}_1^2+k_2 \chi^2 \tilde{\psi}_2 \tilde{\psi}_1^4.
\end{equation}
Here, we rescaled $\tilde \psi_2 \to b^7 \tilde \psi_2$ so that the $c$ dependence of the correlator (\ref{eq:3p2}) is simply~$c^4$, with a similar rescaling also for (\ref{eq:3p1}). The constants $k_i$ can be read off directly from the correlators. Notice that $\mathcal{W}_0$ gives a mass term for $\chi$ when $\tilde \psi_1, \tilde \psi_2 \neq 0$.
Let us finally compute correlators with insertions of $\tau$, which yields the following non-vanishing three-point function:\footnote{Here, $\tilde \psi_1 = \frac{\partial_c Q - \partial_{c'} Q}{c-c'}$, by a slight abuse of notation; note that we cannot use the expression with the gauge equivalent form $\hat Q$, since $\tau$ lives precisely at the points, where (\ref{eq:Q_bound_gauge}) is not well defined.}
\begin{equation}\label{eq:3ptau}
 \langle \tau \tau \tilde \psi_1 \rangle = -\frac{\eta^3 (3 b'^3+b b'^2-b^2 b' -3b^3)}{16 a^3 c^4}.
\end{equation}
Since this correlator is non-vanishing for every point where $\tau$ exists, we conclude that $\tau$ will not give rise to an unobstructed direction in the moduli space.

\begin{figure}
\centering
\includegraphics[scale=0.5]{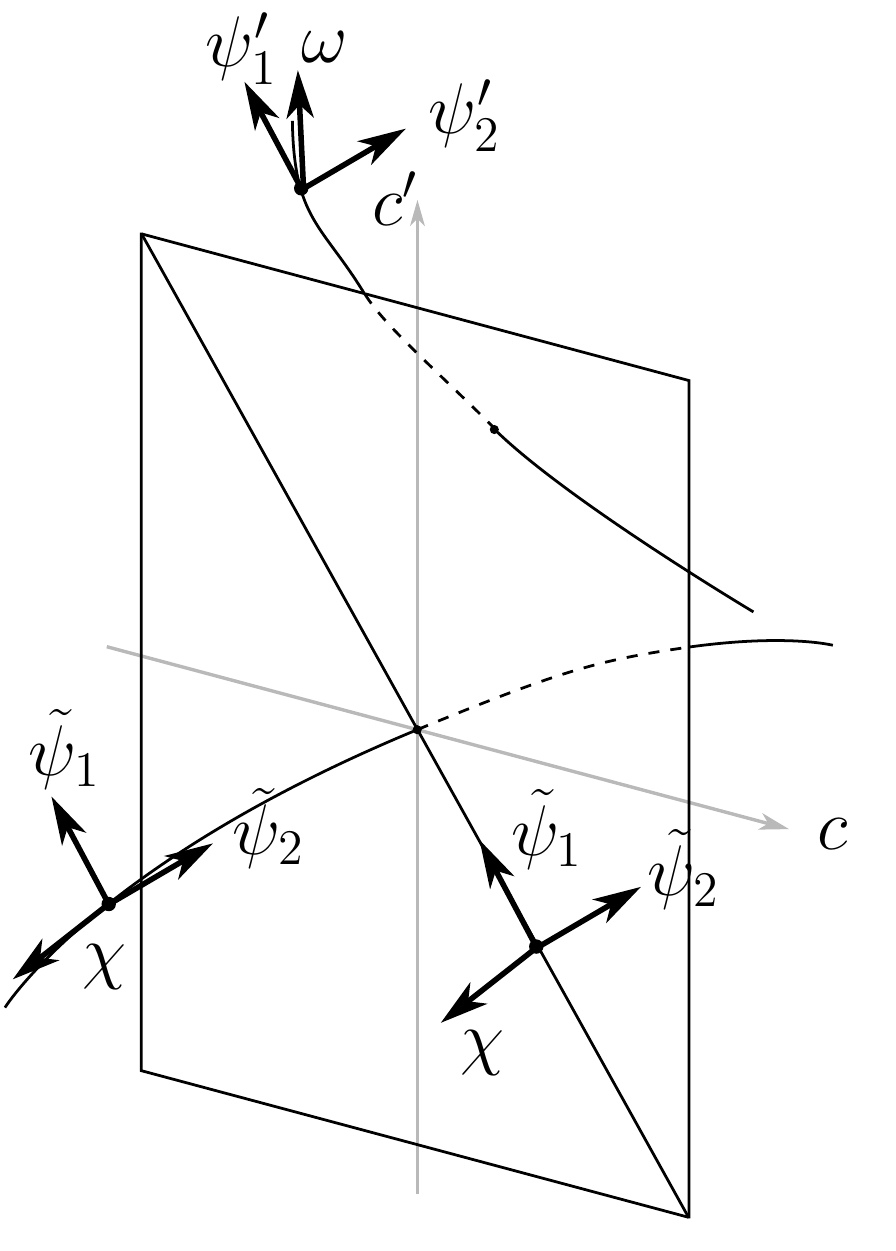}
\caption{The intersections of branch II* with other branches, together with the marginal fermions living on these branches.}
\label{fig:sketch2}
\end{figure}

In summary, we have found that the two-dimensional family of bound states (\ref{eq:Q_bound_eta}) intersects with two other branches at certain special points. The first type of intersection is with branch I, which we found at the points $(a=a',b=b',c=c'=0)$. A similar computation in the patch $c=c'\neq 0$ reveals an additional intersection with a branch of type I at $(a=a'=0,b=b',c=c')$. The second type of intersection is with branch III*, which we found at the points
$(a=a', b=b'=0, c\neq c')$. An interesting feature that we have observed is that the generators of branch II* extend continuously to these other branches, and become obstructed there. On the other hand, we have found that the generators of branch I and branch III* cannot be extended to the entire of branch II*. The situation is illustrated in figure~\ref{fig:sketch2}.

\subsubsection{Branch III}

The last type of branch that one can obtain with the ansatz (\ref{eq:ansatzcurve}) describes curves of the following form:
\begin{align}\label{eq:BranchIII_curves}
(x_1,x_2,x_3,x_4,x_5)&=(\eta u, u, a v^2, b v^2, c v^2), \qquad (u,v)\in \mathbb{P}^1 \;,\\
a^4&+b^4+c^4 =0, \qquad \qquad \eta^8=-1 \ .
\end{align}
The associated matrix factorization is
\begin{align}\label{eq:mfbranchIII}
\begin{array}{ll}
J_1 = x_1 -\eta x_2 &  \qquad \qquad \qquad E_1 =\sum_{n=0}^{7}(\eta x_2)^n x_1^{7-n}\\
J_2 = a x_4 - b x_3 & \qquad \qquad \qquad E_2 = \sum_{n=0}^{3}\frac{(a x_4)^n (b x_3)^{3-n}}{a^4}  \\
J_3 = c x_3 - a x_5  & \qquad \qquad \qquad E_3 = \sum_{n=0}^{3}\frac{(c x_3)^n (a x_5)^{3-n}}{-a^4}.
\end{array}
\end{align}
The marginal spectrum on this branch consists of the modulus $\rho=\partial_b Q$, and an additional fermion $\theta=\frac{x_1^2}{x_3} \partial_b Q$. The three point function
\begin{equation} \label{eq:corrIII}
 \langle \theta \, \theta \, \theta \rangle = -\frac{3 a^2 b^2 \eta^7}{16 c^7}
\end{equation}
reveals that $\theta$ is obstructed at first order, except for the points where $a$, $b$, or $c$ are zero. One can easily see that these are precisely the points of intersection with branches of type II, where $\theta$ is extended to the modulus of the corresponding branch. Note that from (\ref{eq:corrIII}) one can again deduce that the lowest order term in the boundary superpotential~$\mathcal{W}(\theta,\rho)$~is
\begin{equation}
\mathcal{W}_0=\theta^3 \rho^2.
\end{equation}
From this one can also see that $\rho$ becomes massive for $\theta \neq 0$ which is in accord with the fact that it does nt exist on branch II away from the intersection points.

\subsubsection{Branch III*}

As we already mentioned in the discussion of branch II*, there is one additional family of D-branes that we have to include into our analysis. It is given by
\begin{align}\label{eq:mfbranchIII*}
\begin{array}{ll}
J_1 = (x_1 - \eta' x_2)(x_1 - \eta'' x_2) &  \qquad \qquad \qquad E_1 = \prod_{\eta^8=-1, \eta \neq \eta',\eta''}(x_1-\eta x_2)\\
J_2 = a x_4 - b x_3 & \qquad \qquad \qquad E_2 = \sum_{n=0}^{3}\frac{(a x_4)^n (b x_3)^{3-n}}{a^4} \\
J_3 = c x_3-a x_5 & \qquad \qquad \qquad E_3 = \sum_{n=0}^{3}\frac{(c x_3)^n (a x_5)^{3-n}}{-a^4}\\
&\\
\multicolumn{2}{c}{a^4+b^4+c^4 =0, \qquad \qquad \eta'^{8}=\eta''^{8}=-1, \ \eta'\neq \eta''\;.}
\end{array}
\end{align}
Using again the decomposition $Q=Q_1(x_1,x_2) \otimes Q_2 (x_3,x_4,x_5)$, we see that the $Q_2$-part is the same as for branch III, while the $Q_1$-part is a bound state of two permutation branes given by $J_{1A}=x_1 - \eta' x_2$ and $J_{1B}=x_1 - \eta'' x_2$, as was shown in \cite{Brunner:2005fv}. Thus, one can view (\ref{eq:mfbranchIII*}) as a bound state of two D-branes $Q_A$ and $Q_B$ in branch III (taking $Q_{2A} = Q_{2B}$, and $Q_{1A}\neq Q_{1B}$), with the tachyon being of the form $T_f \otimes 1$ ($T_f$ is unique in $H(Q_{1A},Q_{1B})$). Note that since this time we take the boundary moduli of the constituent D-branes to be equal, the resulting family is only one-dimensional. 

There are three marginal fermions in the boundary preserving spectrum on this branch. Apart from the modulus $\omega=\partial_b Q$, one can find $\psi_1'= \frac{x_1^2}{x_3}\partial_b Q$ and $\psi_2'= \frac{x_1 x_2}{x_3}\partial_b Q$. The non-vanishing three-point functions for these fields are\footnote{These expressions can be derived easily from (\ref{eq:corrIII}) and the independence of the correlators on the tachyon configuration, discussed in section \ref{MF:boundstates}.}
\begin{align}\label{eq:corrBranchIII*1}
 \langle \psi_1' \psi_1' \psi_1' \rangle = -\frac{3 a^2 b^2}{16 c^7} (\eta'^7+\eta''^7), \qquad 
 \langle \psi_1' \psi_1' \psi_2' \rangle = -\frac{3 a^2 b^2}{16 c^7} (\eta'^6+\eta''^6),\\ \label{eq:corrBranchIII*2}
 \langle \psi_1' \psi_2' \psi_2' \rangle = -\frac{3 a^2 b^2}{16 c^7} (\eta'^5+\eta''^5), \qquad
 \langle \psi_2' \psi_2' \psi_2' \rangle = -\frac{3 a^2 b^2}{16 c^7} (\eta'^4+\eta''^4).
\end{align}
This shows that both $\psi_1'$ and $\psi_2'$ are obstructed at first order, except for the points where $a$, $b$, or $c$ are zero, which are precisely the points of intersection with branches of type~II*. The lowest order term in the boundary superpotential $\mathcal{W}(\psi_1',\psi_2',\omega)$ can be seen to have the form:
\begin{equation}
\mathcal{W}_0= \omega^2 (k_1 \psi_1'^3+ k_2 \psi_1'^2 \psi_2'+ k_3 \psi_1' \psi_2'^2+ k_4\psi_2'^3),
\end{equation}
which again gives a mass term for $\omega$ when $\psi_1'\neq 0$, or $\psi_2'\neq 0$.

\subsection{Summary: The Heterogeneous Moduli Web of the Model \texorpdfstring{$\mathbb{P}^4_{(11222)}[8]$}{P11222[8]}}

\label{sec:moduliweb}

After the detailed discussion of the various branches, their marginal spectra and intersections,
let us now summarize our findings and comment on the global structure of the moduli space. The starting point of our discussion were one-dimensional families of D2-branes wrapping the rational curves (\ref{eq:ansatzcurve}). Studying these families at the Landau-Ginzburg point, and focusing on the possible boundary perturbations, we have found new types of families that intersect with (\ref{eq:ansatzcurve}) at certain special points. These additional families are 1d- or 2d-branches, and we have shown that they have an interpretation as bound states of the D-branes associated to (\ref{eq:ansatzcurve}). The global structure of the resulting moduli space can then be seen in table (\ref{tab:intersections}), where we list all the intersections between the branches. In particular, we find that the moduli space consists of two disconnected components $\mathcal{M}_1$ and $\mathcal{M}_2$ --- the former comprising the branches of type II and III, and the latter connecting the branches  I, II* and III*. These two components are distinguished by the associated K-theory classes. 

The charges and intersection matrices for the D-branes considered can be computed explicitly at the Landau-Ginzburg point \cite{Walcher:2004tx, Brunner:2005fv, Caviezel:2005th}. 
The data specifying the D-brane does not merely consist of a matrix factorization, but also of a $\ZZ_8$ representation label, such that there are altogether $8$ copies of each family of factorizations, distinguished by a representation label.
The 6-dimensional charge lattice is generated by the $8$ branes whose matrix factorizations constitute $\mathcal{M}_1$; we denote their charges by $e_\varphi$, with $\varphi$ being the $\ZZ_8$ representation label of the brane. This has been checked in \cite{Brunner:2005fv, Caviezel:2005th}, where it was shown that the intersection matrix of these $8$ fractional branes contains a submatrix of maximal rank and determinant 1. 

The branes in $\mathcal{M}_2$ then carry charges $e_\varphi+e_{\varphi+1}$; this is clear, since on branch~II* $\subset  \mathcal{M}_2$ the branes are constructed as bound states of branes in $\mathcal{M}_1$ with subsequent representation labels. 

\medskip
\medskip

The large volume charges can be determined by computing the intersections with the branes given by the matrix factorizations
\begin{equation}
Q= \sum_{i=1}^5 \left( x_i \pi^i + \frac{\partial W}{\partial x_i} \bar{\pi}^i\right)
\end{equation}
(corresponding to the Recknagel-Schomerus branes in conformal field theory)
whose large volume interpretation are the D6-brane and its images under the Landau-Ginzburg monodromy. Explicitly, the charges in the large volume basis have been listed in \cite{Diaconescu:2000ec}. From this, one obtains the following charges for the $\mathcal{M}_1$-branes \cite{Brunner:2005fv}:
\begin{eqnarray}
{\rm ch} (V_1) &=&  -1+h+H-l -L-\frac{5}{3} v\\
{\rm ch} (V_2) &=&  -3h +L\\
{\rm ch} (V_3) &=& 2-3h-H-l+L+\frac{2}{3} v \\
{\rm ch} (V_4) &=& h-L+v \\
{\rm ch} (V_5) &=&  -1+h+l+v\\
{\rm ch} (V_6) &=&  h\\
{\rm ch} (V_7) &=&  h+l \\
{\rm ch} (V_8) &=&  h-v,
\end{eqnarray}
where $v=H^2L$ is a point.
In particular, one finds that at any point in $\mathcal{M}_1$, one of the 8 fractional D-branes carries charge $h$, and its monodromy image has charge $h+l$. The bound states in branch $\mathcal{M}_2$ therefore have charge $2h+l$. This confirms the geometric expectation discussed in section \ref{sec:model}.

\medskip
\medskip

Similarly as in \cite{Baumgartl:2008qp}, we can represent the data from table \ref{tab:intersections} in a graph, by assigning a vertex to each family and an edge to an intersection. Since this graph is unoriented and contains self-intersections, we construct its universal cover, which we display in figure~\ref{fig:mod_space} (see~\cite{Baumgartl:2008qp} for a related discussion).

\begin{table}
\begin{center}
\begin{tabular}{|c|c|c|c|}
\hline
Type & Name & Branch & Intersections \\
\hline
\multirow{6}{*}{I}& $(\gamma)	$&$(14)(325)$&$(\delta), (\nu), (\beta)^*$\\
& $(\delta)	$&$(23)(415)$&$(\gamma), (\epsilon), (\rho)^*$\\
& $(\epsilon)	$&$(15)(423)$&$(\delta), (\mu), (\zeta)^*$\\
& $(\lambda)	$&$(13)(425)$&$(\mu), (\nu), (\rho)^*$\\
& $(\mu)	$&$(24)(315)$&$(\lambda), (\epsilon), (\beta)^*$\\
& $(\nu)	$&$(25)(314)$&$(\gamma), (\lambda), (\zeta)^*$\\
\hline
\multirow{3}{*}{II}&$(\beta)	$&$(35)(214)$&$(\alpha)$\\
& $(\zeta)	$&$(34)(215)$&$(\alpha)$\\
& $(\rho)	$&$(45)(213)$&$(\alpha)$\\
\hline
III& $(\alpha)	$&$(12)(435)$&$(\beta), (\zeta), (\rho)$\\
\hline
\multirow{3}{*}{II*} & $(\beta)^*$ & $(35)(214)^*$ & $(\gamma), (\mu), (\alpha)^*$\\
& $(\zeta)^*	$&$(34)(215)^*$&$(\epsilon), (\nu), (\alpha)^*$\\
& $(\rho)^*	$&$(45)(213)^*$&$(\delta), (\lambda), (\alpha)^*$\\
\hline
III*& $(\alpha)^*$ &$(12)(435)^*$&$(\beta)^*, (\zeta)^*, (\rho)^*$\\
\hline
\end{tabular}
\end{center}
\caption{Intersections between the various branches. The notation $(i_1,i_2)(i_3,i_4,i_5)$ reflects the separation of variables in (\ref{eq:ansatzcurve}).}
\label{tab:intersections}
\end{table}

\begin{figure}
\centering
\begin{tikzpicture}[scale=1.3]
\tikzstyle{every node}=[circle, draw, inner sep = 0pt, minimum width=11pt]

\node (1) at (0,1) {\footnotesize $\rho$};
\node (2) at (-0.866,-0.5)  {\footnotesize $\zeta$};
\node (3) at (0.866,-0.5)  {\footnotesize $\beta$};
\node (4) at (0,0) {\footnotesize $\alpha$};

\draw (1) -- (4);
\draw (2) -- (4);
\draw (3) -- (4);

\end{tikzpicture}
\centering
\begin{tikzpicture}[scale=1.22,star/.style={label={above right,xshift=-2pt, yshift=-2pt}:\footnotesize$*$}]
\tikzstyle{every node}=[circle, draw, inner sep = 0pt, minimum width=11pt]

\node (1) at (1.534, 3.724) {\footnotesize $\delta$};
\node (2) at (-0.948, -3.915)  {\footnotesize $\nu$};
\node (3) at (0.34, 1.402)  {\footnotesize $\nu$};
\node (4) at (-0.549, -1.334)  {\footnotesize $\delta$};
\node (5) at (0.268, 0.369)  {\footnotesize $\mu$};
\node (6) at (0.864, 1.924) {$\;\;\;\;\,$};
\node (6) at (0.864, 1.924) [star] {\footnotesize $\zeta$};
\node (7) at (0., -0.456) [star] {\footnotesize $\alpha$};
\node (8) at (-0.432, -2.065)  {\footnotesize $\epsilon$};
\node (9) at (0., 6.165)  [star] {\footnotesize $\alpha$};
\node (10) at (0.593, 2.836)  {\footnotesize $\epsilon$};
\node (11) at (-3.623, -4.988)  {\footnotesize $\mu$};
\node (12) at (-1.187, -2.644)  {$\;\;\;\;\,$};
\node (12) at (-1.187, -2.644)  [star] {\footnotesize $\zeta$};
\node (13) at (-0.269, 0.37)  {\footnotesize $\epsilon$};
\node (14) at (-0.435, -0.142)  {$\;\;\;\;\,$};
\node (14) at (-0.435, -0.142)  [star] {\footnotesize $\zeta$};
\node (15) at (-0.594, 2.836)  {\footnotesize $\mu$};
\node (16) at (-2.147, -1.946)  [star] {\footnotesize $\alpha$};
\node (17) at (4.016, -0.308)  {\footnotesize $\lambda$};
\node (18) at (3.431, -2.111)  {$\;\;\;\;\,$};
\node (18) at (3.431, -2.111)  [star] {\footnotesize $\rho$};
\node (19) at (4.863, 1.581)  {\footnotesize $\nu$};
\node (20) at (3.006, -4.137)  {\footnotesize $\delta$};
\node (21) at (1.229, 0.756)  {$\;\;\;\;\,$};
\node (21) at (1.229, 0.756)  [star] {\footnotesize $\rho$};
\node (22) at (0.55, -1.334)  {\footnotesize $\lambda$};
\node (23) at (5.863, 1.906)  {$\;\;\;\;\,$};
\node (23) at (5.863, 1.906)  [star] {\footnotesize $\zeta$};
\node (24) at (3.623, -4.988)  {\footnotesize $\epsilon$};
\node (25) at (-0.6, 0.826)  {\footnotesize $\delta$};
\node (26) at (-0.971, -0.315)  {\footnotesize $\nu$};
\node (27) at (-1.534, 3.724)  {\footnotesize $\lambda$};
\node (28) at (-3.43, -2.111)  {$\;\;\;\;\,$};
\node (28) at (-3.43, -2.111)  [star] {\footnotesize $\rho$};
\node (29) at (0.97, -0.315)  {\footnotesize $\gamma$};
\node (30) at (-4.864, 1.58)  {\footnotesize $\gamma$};
\node (31) at (1.563, 1.416)  [star] {\footnotesize $\alpha$};
\node (32) at (0.432, -2.065)  {\footnotesize $\mu$};
\node (33) at (1.438, 0.11)  {\footnotesize $\delta$};
\node (34) at (1.099, -0.934)  {\footnotesize $\nu$};
\node (35) at (-1.228, 0.757)  {$\;\;\;\;\,$};
\node (35) at (-1.228, 0.757)  [star] {\footnotesize $\rho$};
\node (36) at (-1.438, 0.11)  {\footnotesize $\lambda$};
\node (37) at (-1.563, 1.416)  [star] {\footnotesize $\alpha$};
\node (38) at (-2.097, -0.227)  {\footnotesize $\mu$};
\node (39) at (-2.514, 1.44)  {$\;\;\;\;\,$};
\node (39) at (-2.514, 1.44)  [star] {\footnotesize $\zeta$};
\node (40) at (-2.88, 0.312)  {\footnotesize $\epsilon$};
\node (41) at (3.068, 2.61)  {\footnotesize $\gamma$};
\node (42) at (0.948, -3.914)  {\footnotesize $\gamma$};
\node (43) at (-0.34, 1.402)  {\footnotesize $\gamma$};
\node (44) at (-1.098, -0.934)  {\footnotesize $\gamma$};
\node (45) at (2.096, -0.227)  {\footnotesize $\epsilon$};
\node (46) at (1.83, -1.049)  {$\;\;\;\;\,$};
\node (46) at (1.83, -1.049)  [star] {\footnotesize $\zeta$};
\node (47) at (2.514, 1.441)  {$\;\;\;\;\,$};
\node (47) at (2.514, 1.441)  [star] {\footnotesize $\beta$};
\node (48) at (1.187, -2.643)  {$\;\;\;\;\,$};
\node (48) at (1.187, -2.643)  [star] {\footnotesize $\beta$};
\node (49) at (-0.864, 1.924)  {$\;\;\;\;\,$};
\node (49) at (-0.864, 1.924)  [star] {\footnotesize $\beta$};
\node (50) at (-1.83, -1.049)  {$\;\;\;\;\,$};
\node (50) at (-1.83, -1.049)  [star] {\footnotesize $\beta$};
\node (51) at (-3.068, 2.61)  {\footnotesize $\nu$};
\node (52) at (-4.016, -0.308)  {\footnotesize $\delta$};
\node (53) at (-5.863, 1.905)  {$\;\;\;\;\,$};
\node (53) at (-5.863, 1.905)  [star] {\footnotesize $\beta$};
\node (54) at (2.88, 0.312)  {\footnotesize $\mu$};
\node (55) at (2.147, -1.946)  [star] {\footnotesize $\alpha$};
\node (56) at (0.435, -0.141)  {$\;\;\;\;\,$};
\node (56) at (0.435, -0.141)  [star] {\footnotesize $\beta$};
\node (57) at (0.599, 0.825)  {\footnotesize $\lambda$};
\node (58) at (0., -1.02)  {$\;\;\;\;\,$};
\node (58) at (0., -1.02)  [star] {\footnotesize $\rho$};
\node (59) at (0., 5.113)  {$\;\;\;\;\,$};
\node (59) at (0., 5.113)  [star] {\footnotesize $\rho$};
\node (60) at (-3.005, -4.137)  {\footnotesize $\lambda$};

\draw (1) -- (10) ;
\draw (1) -- (41) ;
\draw (1) -- (59) ;
\draw (2) -- (12) ;
\draw (2) -- (42) ;
\draw (2) -- (60) ;
\draw (3) -- (6) ;
\draw (3) -- (43) ;
\draw (3) -- (57) ;
\draw (4) -- (8) ;
\draw (4) -- (44) ;
\draw (4) -- (58) ;
\draw (5) -- (13) ;
\draw (5) -- (56) ;
\draw (5) -- (57) ;
\draw (6) -- (10) ;
\draw (6) -- (31) ;
\draw (7) -- (14) ;
\draw (7) -- (56) ;
\draw (7) -- (58) ;
\draw (8) -- (12) ;
\draw (8) -- (32) ;
\draw (9) -- (23) ;
\draw (9) -- (53) ;
\draw (9) -- (59) ;
\draw (10) -- (15) ;
\draw (11) -- (24) ;
\draw (11) -- (53) ;
\draw (11) -- (60) ;
\draw (12) -- (16) ;
\draw (13) -- (14) ;
\draw (13) -- (25) ;
\draw (14) -- (26) ;
\draw (15) -- (27) ;
\draw (15) -- (49) ;
\draw (16) -- (28) ;
\draw (16) -- (50) ;
\draw (17) -- (18) ;
\draw (17) -- (19) ;
\draw (17) -- (54) ;
\draw (18) -- (20) ;
\draw (18) -- (55) ;
\draw (19) -- (23) ;
\draw (19) -- (41) ;
\draw (20) -- (24) ;
\draw (20) -- (42) ;
\draw (21) -- (31) ;
\draw (21) -- (33) ;
\draw (21) -- (57) ;
\draw (22) -- (32) ;
\draw (22) -- (34) ;
\draw (22) -- (58) ;
\draw (23) -- (24) ;
\draw (25) -- (35) ;
\draw (25) -- (43) ;
\draw (26) -- (36) ;
\draw (26) -- (44) ;
\draw (27) -- (51) ;
\draw (27) -- (59) ;
\draw (28) -- (52) ;
\draw (28) -- (60) ;
\draw (29) -- (33) ;
\draw (29) -- (34) ;
\draw (29) -- (56) ;
\draw (30) -- (51) ;
\draw (30) -- (52) ;
\draw (30) -- (53) ;
\draw (31) -- (47) ;
\draw (32) -- (48) ;
\draw (33) -- (45) ;
\draw (34) -- (46) ;
\draw (35) -- (36) ;
\draw (35) -- (37) ;
\draw (36) -- (38) ;
\draw (37) -- (39) ;
\draw (37) -- (49) ;
\draw (38) -- (40) ;
\draw (38) -- (50) ;
\draw (39) -- (40) ;
\draw (39) -- (51) ;
\draw (40) -- (52) ;
\draw (41) -- (47) ;
\draw (42) -- (48) ;
\draw (43) -- (49) ;
\draw (44) -- (50) ;
\draw (45) -- (46) ;
\draw (45) -- (54) ;
\draw (46) -- (55) ;
\draw (47) -- (54) ;
\draw (48) -- (55) ;

\end{tikzpicture}
\caption{The two components of the moduli space. In this diagram, the vertices correspond to the moduli branches (listed in table \ref{tab:intersections}), and the edges represent intersections. The circular vertices correspond to one-dimensional branches, while the double circles indicate a two-dimensional branch. Vertices with the same label are identified.}
\label{fig:mod_space}
\end{figure}

\subsection{Moduli Web for the Quintic}

The moduli space of D2-branes wrapping lines on the quintic was studied in~\cite{Baumgartl:2007an,Baumgartl:2008qp}. In this section we wish to construct bound states of such D-branes and analyze the corresponding moduli space. As we will see, this moduli space, which is decoupled from the moduli space of lines, consists of 1d- and 2d-branches of bound states, and exhibits a particularly simple structure.

Let us start  by briefly summarizing the results of~\cite{Baumgartl:2007an,Baumgartl:2008qp}. At the Gepner point the quintic is described by the LG superpotential
\eqn{
	W=x_1^5+x_2^5+x_3^5+x_4^5+x_5^5 \ .
}	
The matrix factorizations associated to the lines
$(x_1,x_2,x_3,x_4,x_5)=(\eta u, u, a v, b v, c v)$,  $(u,v)\in \mathbb{P}^1$,
are given by the set of polynomials
\eqn{
\label{quintic:J}
	J_1 = x_1-\eta x_2, \quad J_2 = ax_4-bx_3, \quad J_3 = cx_3-ax_5\ .
}
The parameters appearing in these expressions must satisfy
\eqn{
	{\cal M}:\; a^5+b^5+c^5 = 0\ , \qquad {\cal N}:\; \eta^5=-1\ ,
}
in order to describe supersymmetric solutions. 
Similarly as in the example $\PP^4_{(11222)}[8]$, the moduli space of (\ref{quintic:J}) is the product of the Riemann curve ${\cal M}$ and the discrete set ${\cal N}$ of the fifth roots of $-1$, and hence defines a complex one-dimensional branch.
Other branches can be found by using the symmetries of the quintic. They are obtained simply by acting with the permutation group on the coordinates in the defining polynomials (\ref{quintic:J}), which gives a moduli web of 1d-branches \cite{Baumgartl:2007an}.

Along the branch (\ref{quintic:J}) the full odd cohomology is obtained by multiplying the derivative of $Q$ with suitable rational functions.
We find two marginal fermionic fields in the spectrum
\eqn{
	\frac{\partial}{\partial b} Q,\qquad\text{and}\qquad\frac{x_1}{x_3}\frac{\partial}{\partial b} Q\ .
}
The first fermion generates the 1d-branch and therefore is necessarily exactly marginal. The second fermion is generically obstructed~\cite{Baumgartl:2007an}, which can for example be seen from its non-vanishing three-point function
\eqn{\label{eq:3ptquintic}
\Big\langle \ \Big(\frac{x_1}{x_3}\frac{\partial}{\partial b} Q\Big)^3 \ \Big\rangle = -\frac{2}{5} \eta^4 \frac{b^3}{c^9}\ .
}
The 1d-branches in the moduli web intersect in points, where the roles of the obstructed and unobstructed fermions are interchanged.

We now construct bound states of the form (\ref{eq:boundtensor}) from the D-branes just described. For this, we again use the tensor decomposition $Q=Q_1\otimes Q_2$, where $Q_1$ is given by the pair $(J_1,E_1)$ depending only on the variables $(x_1,x_2)$, and $Q_2$ is given by $(J_2,E_2)$, $(J_3,E_3)$ depending on $(x_3,x_4,x_5)$.

Let us first consider the case where the bound state is formed out of $Q_{1A}(\eta)\otimes Q_2(a,b,c)$ and $Q_{1B}(\eta')\otimes Q_2(a,b,c)$, and where the boundary changing fermion is of the form $T \otimes 1$ with $T \in H(Q_{1A}, Q_{1B})$. Switching on this fermion starts a condensation process which is independent of the $Q_2$-part, so it can be  factored out. The result of the condensation of $Q_{1A}(\eta)\oplus Q_{1B}(\eta')$ gives \cite{Brunner:2005fv}:
\eqn{
	Q_1(\eta,\eta') = \begin{pmatrix}0 & J_1\\ E_1 & 0\end{pmatrix},
}
with
\eqn{
	 J_1 = (x_1-\eta x_2)(x_1-\eta' x_2),\qquad\text{and}
	\qquad
	 E_2 = \prod_{\eta''^5=-1,\eta''\ne\eta,\eta'}(x_1-\eta'' x_2)\ .
}
Thus the bound state, depending now on one continuous modulus and two discrete moduli $\eta$ and $\eta'$, is
\eqn{
\label{bound:t1}
	 Q(\eta,\eta';a,b,c) =  Q_1(\eta,\eta') \otimes Q_2(a,b,c)\ .
}
The spectrum contains one exactly marginal field
\eqn{
\label{MF:1branch_exact}
	\frac{\partial}{\partial b} Q
}
which generates translations on this 1d-branch. There are two more marginal fermions in the spectrum with representatives
\eqn{
\label{MF:1branch_obs}
	\frac{x_1}{x_3}\frac{\partial}{\partial b} Q \ ,
		\qquad\text{and}\qquad
	\frac{x_2}{x_3}\frac{\partial}{\partial b} Q \ .
}
Their three-point functions\footnote{Similarly as in (\ref{eq:corrBranchIII*1}),(\ref{eq:corrBranchIII*2}), these three-point functions can be easily computed using (\ref{eq:3ptquintic}) and~(\ref{corrsum}).} show that away from the points where $a$, $b$, or $c$ are zero, these two fermions are obstructed at first order.

Next, we turn to bound states which share the same $Q_1$ and condense two copies of $Q_2$ at different points of $\mathcal{M}$, which we denote by $Q_{2A}(a,b,c)$ and $Q_{2B}(a',b',c')$. Again, we choose a boundary changing fermion of the form $1 \otimes T$ with $T\in H(Q_{2A}, Q_{2B})$, so we can factor out the $Q_1$-part.
As in section \ref{sec:BranchII*}, a representative for $T$ is given by the expression~(\ref{Tachyon}).
The bound state is then given by a matrix factorization $Q(\eta; a, b, c, a', b', c')$ with a two-dimensional moduli space. There are obviously two exactly marginal fields in the spectrum, given by
\eqn{
\label{MF:2branch_exact}
	\frac{\partial}{\partial c}  Q \qquad\text{and}\qquad \frac{\partial}{\partial c'}  Q \ ,
}
spanning the moduli space ${\cal N}\times{\cal M}\times{\cal M}$. One can find three additional marginal fermions in the spectrum. Two of them live only at the points $(a=a', b=b', c=c')$ with $a$, $b$, or $c$ equal to zero, where they are unobstructed. The third fermion $\chi$ can be found everywhere on the 2d-branch, and has a representative given by the linear combination
\eqn{
\label{MF:2branch_obs}
	\chi=\frac{x_1}{x_3}\left(\frac{\partial}{\partial c} Q-\mu \frac{\partial}{\partial c'} Q\right) \ ,
}
where $\mu$ is defined as
\eqn{
	\mu = \frac{b'^4}{b^4} \;\frac{1 + b^4 b' + c^4 c'}{1 + b'^4 b + c'^4 c} \ .
} 
Its 3-point function can be computed using (\ref{eq:3ptquintic}) and (\ref{corrsum}) as 
\eqn{
\langle \chi \chi \chi \rangle = -\frac{2}{5} \eta^4 \left(\frac{c^3}{b^9} - \mu^3 \frac{c'^3}{b'^9}\right).
}
Note that this correlator vanishes at the points $(a=a', b=b', c=c')$, but this is just due to the fact that the representative (\ref{MF:2branch_obs}) itself is zero at these points; choosing another representative for $\chi$ here, one can again find an obstruction at first order. We conclude that $\chi$ is obstructed everywhere on the 2d-branch except for the points $(a=a', b\neq b', c=c'=0)$, and similarly for $a=a'=0$, and $b=b'=0$.

Let us now inspect the intersections between the 1d- and 2d-branches of bound states we have just constructed.
First, we consider the point  $(a=a', b\neq b', c=c'=0)$ on the 2d-branch.
Taking $a=1$, the matrix factorization in this limit becomes\footnote{Here one can again use the gauge equivalent form given in Appendix \ref{sec:App_matrices}.}
\eqn{
\label{quintic:2intersection}
	J_1=x_1-\eta x_2\ ,\qquad
	J_2=(x_4-b x_3)(x_4-b' x_3)\ , \qquad
	J_3=x_5 \ ,
}
with three marginal fermions in the spectrum. This matrix factorization, though, also lies in 
the one-dimensional family of bound states with $Q_1$ given by $ J_1= (x_4-\eta' x_3)(x_4-\eta'' x_3)$ and $Q_2(a'', b'', c'')$ given by $J_2 = a'' x_1 - b'' x_2$, $J_3 = c'' x_2-a'' x_5$. 
In the limit $c''\to0$ with $a''=1$, the matrix factorization becomes identical to (\ref{quintic:2intersection}) as long as $b''=\eta$, $b=\eta'$ and $b'=\eta''$, hence the two branches intersect.
Note also that  the marginal cohomologies are mapped into each other at the intersection point. The three fermions coming from the 1d-branch can be identified directly with the three fermions coming from the 2d-branch, where the role of obstructed and exactly marginal fermions is interchanged.

The other intersection point can be found on the 2d-branch for $(a=a', b=b', c=c')$ with $b=0$ (and similarly for $a=0$, or $c=0$). For $a=1$ and $c=\eta'$, the matrix factorization at this point becomes
\eqn{
	J_1=x_1-\eta x_2\ ,\qquad
	J_2=x_4^2\ , \qquad
	J_3=x_5-\eta' x_3\ ,
}
which by exchanging $J_1$ and $J_3$ clearly defines a matrix factorization in another 2d-branch. The marginal spectrum at this point contains five fermions --- four of them correspond simply to the generators of the two 2d-branches intersecting at this point, and the remaining one can be identified with the obstructed fermion $\chi$. Note that in this case the generators of the 2d-branch do not extend to the other 2d-branch.

To summarize, for each single D2-brane branch of the form (\ref{quintic:J}) we have constructed two kinds of bound state branches --- one 1d-branch and one 2d-branch. In general, these bound state branches intersect in points, with 1d-2d and 2d-2d intersections being possible. One can easily see that an intersection occurs if and only if the corresponding single D2-brane branches intersect. Let us note that unlike the  $\PP^4_{(11222)}[8]$ case, it is not possible to transit from the moduli web of the single D2-branes to the moduli web of bound states.  A feature that is shared by the quintic and the $\PP^4_{(11222)}[8]$ model is that the moduli web of bound states contains both one- and two-dimensional branches, where the dimension jumps at joints connecting them.

\section{Effective Superpotentials}
\label{sec:Bulk-deformations}

\subsection{Bulk Perturbations from the Untwisted Sector}

When one turns on bulk fields from the $(c,c)$-ring (corresponding to complex structure deformations), one expects that the brane moduli space gets lifted by superpotential terms. Generically, it will collapse to a set of discrete points.

Perturbations from the untwisted sector of a boundary Landau-Ginzburg model are given as monomials perturbing the LG superpotential. We will denote the perturbing polynomials as $G$. 
We can choose the polynomials so that the open string operator $\partial_b Q$ generating the branch becomes obstructed by a superpotential. This is reflected in a non-vanishing correlator
\begin{equation}\label{eq:Bbulkboundary}
B_{G \partial_b Q} = \langle G \partial_b Q \rangle
\end{equation}
that can be then integrated to obtain a superpotential term~\cite{Baumgartl:2007an,Baumgartl:2008qp,Baumgartl:2010ad}. Since $B$ can be calculated at any point on the branch, the result is exact in the boundary couplings. It turns out that $B$ is always a holomorphic one-form on the curve that describes the moduli space of the brane.

In geometric language, the perturbation corresponds to adding a term of degree 2 to the defining polynomial
\begin{equation}
x_1^8+x_2^8+x_3^4+x_4^4+x_5^4 + G(x_1, \dots, x_5)=0 \ .
\end{equation}
In general, families of solutions $J_i=0$ of the undeformed equation will no longer fulfill the perturbed equation. Rather, inserting $J_i=0$ into the perturbed equation will give an additional condition on the parameters. The solutions of this equation describe the supersymmetric brane vacua after the perturbation, that is, the minima of the superpotential. In geometry, techniques to compute the superpotential in terms of relative periods have been developed in \cite{Aganagic:2001nx,Lerche:2002yw,Walcher:2006rs,Jockers:2008pe,Grimm:2008dq,Alim:2009rf}. The equivalence of the geometric and LG approach has been demonstrated (to first order in the bulk) in \cite{Aganagic:2009jq,Baumgartl:2010ad}.

We now wish to compute the correlators (\ref{eq:Bbulkboundary}) explicitly for the branches studied in section \ref{sec:branches}, and integrate them to obtain terms in the effective superpotential. As we will see, the superpotentials are always given in terms of hypergeometric functions that arise as chain integrals on the moduli space. In the case that the brane is a bound state, the superpotential will be additive, i.e., a sum of the superpotentials on the constituent branes, as explained in section \ref{MF:boundstates}. The results are summarized in table \ref{tab:Wbulk}.

\begin{landscape}
\begin{table}
\centering \setlength{\tabcolsep}{5pt} \renewcommand{\arraystretch}{1.75}
\begin{tabular}{|c|c|c|c|}
\hline
Branch & $G$ & $B_{G \partial Q}$ & $\mathcal{W}$ \\ \hline
I & $s^{(5)}(x_1,x_4) s^{(3)}(x_3,x_2,x_5)$ & $\langle G \partial_b Q \rangle = \frac{a \eta s^{(5)}(\eta,1) s^{(3)}(a,b,c)}{8 c^3}$ &
$ \mathcal{W}^{\text{I}}(1,b,c)=\frac{\eta s^{(5)}(\eta,1)}{8} \displaystyle \sum_{q,r,s}\frac{s^{(3)}_{qrs}(-b)^{r+1}}{r+1} \, _{2}F_{1}(\tfrac{r+1}{8},\tfrac{3-s}{4};\tfrac{9+r}{8};b^8)$\\ \hline

II & $s^{(4)}(x_3,x_5) s^{(4)}(x_1,x_4,x_2)$ & $\langle G \partial_c Q \rangle =  \frac{a \eta s^{(4)}(\eta,1) s^{(4)}(a,b,c)}{16 b^3}$ &
$\mathcal{W}^{\text{II}}(1,b,c)=\frac{\eta s^{(4)}(\eta,1)}{16} \displaystyle \sum_{q,r,s} \frac{s^{(4)}_{qrs}(-c)^{r+1}}{r+1} \, _{2}F_{1}(\tfrac{r+1}{8},\tfrac{3-s}{4};\tfrac{9+r}{8};c^8)$\\ \hline

III & $s^{(6)}(x_1,x_2) s^{(2)}(x_3,x_4,x_5)$ & $\langle G \partial_b Q \rangle = \frac{a \eta s^{(6)}(\eta,1) s^{(2)}(a,b,c)}{-32 c^3}$ & $\mathcal{W}^{\text{III}}(1,b,c;\eta)=\frac{\eta s^{(6)}(\eta,1)}{-32} \displaystyle \sum_{q,r,s} \frac{s^{(2)}_{qrs} (-b)^{r+1}}{r+1} \, _{2}F_{1}(\tfrac{r+1}{4},\tfrac{3-s}{4};\tfrac{5+r}{4};b^4)$ \\ \hline

\multirow{2}{*}{II*} & \multirow{2}{*}{$s^{(4)}(x_3,x_5) s^{(4)}(x_1,x_4,x_2)$} & $\langle G \partial_c Q \rangle =  \frac{a \eta s^{(4)}(\eta,1) s^{(4)}(a,b,c)}{16 b^3}$ &
\multirow{2}{*}{$\mathcal{W}^{\text{II*}}(1,b,c,1,b',c')=\mathcal{W}^{\text{II}}(1,b,c)+\mathcal{W}^{\text{II}}(1,b',c')$}\\

& & $\langle G \partial_{c'} Q \rangle =  \frac{a' \eta s^{(4)}(\eta,1) s^{(4)}(a',b',c')}{16 b'^3}$ & \\
\hline

\multirow{2}{*}{III*} & \multirow{2}{*}{$s^{(6)}(x_1,x_2) s^{(2)}(x_3,x_4,x_5)$} & $\langle G \partial_b Q \rangle =   $ & \multirow{2}{*}{$\mathcal{W}^{\text{III*}}(1,b,c;\eta',\eta'')=\mathcal{W}^{\text{III}}(1,b,c;\eta')+\mathcal{W}^{\text{III}}(1,b,c;\eta'')$} \\

& & $\frac{a s^{(2)}(a,b,c) \sum_{\eta \in \{\eta', \eta''\}}\eta s^{(6)}(\eta,1)}{-32 c^3}$ & \\ \hline
\end{tabular}
\caption{List of branches together with the superpotentials $\cal W$ obtained by integrating $B_{G \partial Q}$ over the moduli space. Here, we always choose $G$ to be of the form $G=s^{(d_1)} (x_{i_1}, \; x_{i_2}) s^{(d_2)} (x_{i_3}, \; x_{i_4}, \; x_{i_5})$, where $s^{(d)}$ denotes a quasi-homogeneous polynomial of degree $d$. We also use the notation $s^{(d_2)}_{qrs}$ for the coefficient of the term $s^{(d_2)}_{qrs} x_{i_3}^q x_{i_4}^r  x_{i_5}^s$~in~$s^{(d_2)}$.}
\label{tab:Wbulk}
\end{table}
\end{landscape}

\subsection{Twisted Sector Deformations} 

The main example considered in this paper has a $(c,c)$ ring that consists of elements of the untwisted as well as the twisted sector. The untwisted sector deformations are realized in terms of the Jacobi ring of the superpotential, whereas the twisted sector states arise as charge $(q_L,q_R)=(1,1)$ states that exist because some of the fields are invariant under a subgroup of the full orbifold group. 

In geometric language, the untwisted sector states correspond to perturbations of the defining equation by adding additional monomials, whereas the twisted sector moduli correspond to non-toric deformations. These can be handeled by mapping the model to a birationally equivalent model, where the deformations have a toric realization. For the model at hand this has been worked out in \cite{Candelas:1993dm}.

In the context of D-branes, it is of course of interest to determine the superpotential contributions of the twisted sector moduli. These are expected to couple to marginal operators in the open string sector of suitable D-branes, just like in the case of non-twisted moduli.

Let us be more concrete in our example, starting with a geometric description. Because of the projective equivalence of the model, there exists a ${\mathbb Z}_2$ action under which $x_3, x_4, x_5$ are invariant, whereas $(x_1, x_2)$ get mapped to $(-x_1, -x_2)$. Hence, there is a ${\mathbb Z}_2$ singularity along the curve
\begin{equation} \label{eq:singular_curve}
C: \quad x_1=x_2=0, \quad \quad x_3^4+x_4^4+x_5^4=0 \ .
\end{equation}
This genus $g=3$ curve is resolved by replacing every point by an exceptional ${\mathbb P}_1$.  The non-toric deformations can be associated to the $H^{(1,0)}$ forms of the curve. This is accomplished by a map (see \cite{Katz:1996ht} for a discussion in a physics context)
\begin{equation}
\Phi: H^{(1,0)}(C) \to H^{(2,1)} (M) \ .
\end{equation}
On the level of 3-cycles, this map is given by sending the 1-cycle on the curve to the 3-cycle swept out by the fibers of the exceptional divisor.
Since the differential on $C$  has $2g-2$ simple zeros, after perturbation one expects to be left with $2g-2$ isolated holomorphic ${\mathbb P}_1$'s \cite{Kachru:2000an}.

In the language of D-branes, in the unperturbed model there is a family of D2-branes wrapping the exceptional ${\mathbb P}_1$; the moduli space consists of the set (\ref{eq:singular_curve}).
Turning on a non-toric bulk deformation should then generate a superpotential with $2g-2$ minima, as outlined already in \cite{Kachru:2000an}. 

Here, we will give a concrete calculation of the superpotential responsible for this lifting. For this, we use the following description of the D2-branes wrapping the exceptional D2's in terms of matrix factorizations:
\begin{equation}\label{eq:twistbranes}
J_1= x_1, \quad J_2=x_2, \quad J_3= ax_4-bx_3, \quad J_4= c x_3-a x_5 \ .
\end{equation}
The charge of these branes has been calculated in \cite{Brunner:2005fv}, and indeed these branes correspond to D2-branes wrapping the exceptional ${\mathbb P}_1$'s. The moduli space is generated by the derivative of the BRST charge $Q$.
It is not difficult to see that the moduli space is unaffected by perturbations from the untwisted sector, hence, the superpotential is flat in these directions.

Twisted sectors arise since an integer charge projection is performed in the Landau-Ginzburg model, and the underlying conformal field theory needs to be modular invariant. In the twisted sector, the left- and right movers differ by an application of the spectral flow operator. It takes the form ${\cal U}_\theta= \exp (2\pi i \theta J_0)$, where the eigenvalues of $J_0$ are the $U(1)$ charges of the closed string operators, and $\theta$ denotes the amount of spectral flow. For an integer charge projection, $\theta=1$. A recipe how to calculate the twisted sector contributions and their $U(1)$ charge in the Landau-Ginzburg picture has been given in \cite{Vafa:1989xc}, to which we refer for further details.

In our model, there are $3$ $(c,c)$ fields of charge $1$ coming from the twisted sector that we denote by $\phi_i, i=1,2,3$. Their $U(1)$ charges in the $5$ individual minimal model factors of the full theory can be calculated to be
\begin{equation}
\begin{array}{c||c|c|c|c|c}
        &q_1 & q_2 & q_3 & q_4 & q_5 \\
        \hline\hline
\phi_1  & 3/4&3/4  & 1/2 & 0  & 0 \\
\phi_2  & 3/4&3/4  &  0 & 1/2  & 0 \\
\phi_3  & 3/4&3/4  & 0 & 0  & 1/2 
\end{array}
\end{equation}

To calculate the superpotential, we need to determine the bulk-boundary coupling between the twist fields $\phi_i$ and the marginal operator $\partial_b Q$. The Kapustin-Li formula was initially derived for correlation functions in the unorbifolded Landau-Ginzburg model; its extension to the orbifold case is clear as long as all fields involved are projections from the unorbifolded theory. This is not the case for the twist fields. However, since all boundary fields have their origin in the unorbifolded theory, one can make use of the standard Kapustin-Li formula once one knows the image of the twist field under the bulk-boundary map. In our case, this can be determined using $U(1)$-selection rules. There are a priori two types of boundary operators compatible with $U(1)$ charge selection rules. One of them are the images of the untwisted closed sector deformations $x_1^3 x_2^3 x_{j}$, ($j=1,2,3$) which have the same $U(1)$ charges as the twist fields but do not couple to the brane. However, a second candidate arises, since the open string spectrum of the first two factors of the tensor product of matrix factorizations in (\ref{eq:twistbranes}) contain each a fermion $\omega_i$ ($i=1,2$) of charge $3/4$ that can be used to build a bosonic operator of the right charges.
The natural image of the twist fields compatible with charge selection rules is therefore
\begin{equation}
\phi_1 \to \omega_1 \omega_2 x_3, \quad \phi_2 \to \omega_1 \omega_2 x_4, \quad \phi_3 \to \omega_1 \omega_2 x_5 \ .
\end{equation}
Using this map, the Kapustin-Li formula gives the result:
\begin{equation}
\langle \omega_1 \omega_2 s^{(2)}(x_3,x_4,x_5) \; \partial_b Q \rangle = \frac{a s^{(2)}(a,b,c)}{4 c^3},
\end{equation}
which again is a holomorphic one-form on the moduli space. By integration, we obtain the superpotential
\eqn{\mathcal{W}(1,b,c)=\frac{1}{4} \sum_{q,r,s} \frac{1}{r+1} s^{(2)}_{qrs} (-b)^{r+1}  \, _{2}F_{1}(\tfrac{r+1}{4},\tfrac{3-s}{4};\tfrac{5+r}{4};b^4).}
The superpotential shows that after perturbation we remain with 4 supersymmetric vacua. For the $g=3$ curve (\ref{eq:singular_curve}) this is the expected result.

\section{Conclusions}

In this paper we have investigated  the moduli space of D2-branes in the K3-fibration $\mathbb{P}^4_{(11222)}[8]$, using and extending Landau-Ginzburg techniques developed earlier in \cite{Baumgartl:2007an,Baumgartl:2008qp}. There are various types of 2-branes in this model, distinguished by their charges, and we have focused on branes with low charge. On the one hand, we examined D2-branes wrapping rational curves obtained by embedding $\PP^1$s by maps of homogeneous degree one, on the other hand we considered D2-branes wrapping the $\PP^1$ originating from the resolution of the singularity of the ambient space.

The former class of branes falls into two different classes of different charges, so that the moduli space consists of two disconnected patches. In the first patch, the branes carry a single unit of D2-brane charge, and the moduli space consists of a web of different branches. The dimension of the moduli space is one on all of these branches. In the second patch, the branes carry higher charge: indeed, on some branches in this patch we have shown explicitly that the D-brane can be represented in terms of a bound state of two branes of lower charge. This is however not the case on other branches within the same patch. Also, the dimension of the different intersecting moduli branches can be different on the various branches. Branches of different dimensions intersect at special points.

Although the focus has been on the Calabi-Yau $\mathbb{P}^4_{(11222)}[8]$ the methods presented can be straightforwardly applied to other models.

We also considered the moduli web of D2-branes and their bound states for the quintic. In this case, $h^{1,1}=1$ and there is only one type of D2-brane charge; single D2-branes wrapped on the rational curves of degree one carry a single unit of this charge. The moduli space of these branes consists of several intersecting branches, all of which are one-dimensional \cite{Baumgartl:2007an}. In this paper, we constructed bound states of these branes (carrying two units of D2-brane charge) and studied their moduli space. The moduli web of bound states consists of one-dimensional as well as two-dimensional branches that intersect in points. The intersection pattern is very simple and inherited from the constituent branes: bound state branches intersect if the underlying single brane branches intersect. 

In both examples, we have explicitly computed the spectrum of marginal operators on all D2-brane families and shown how they are mapped from one branch to the other at joints between intersecting branches. The branch itself is generated by unobstructed operators, and their number determines the dimension of the branch. Besides that we have found marginal operators that are obstructed and therefore do not span an additional dimension. Roughly speaking, when jumping from one branch to the other the role of the obstructed and unobstructed operators are exchanged. This is in particular also true for joints connecting branches of different dimension. On the 1d-branch we have identified two obstructed marginal operators, which are mapped to the two unobstructed marginal operators of the 2d-branch. Also the single unobstructed operator of the 1d-branch becomes obstructed on the 2d-branch. In this way, we could follow the marginal operators throughout the connected pieces of the moduli space. This moduli space can be regarded as the valley of a superpotential, and depending on the precise location within the valley different operators are truly marginal.

Based on the construction of the moduli web of the two-parameter model, we have computed bulk-induced effective superpotentials which appear under toric as well as non-toric deformations of the background geometry. These superpotentials are exact in the open string moduli and first order in the closed string. In the case of bound states which have two free moduli we have shown that the boundary changing tachyon does not generate a superpotential, so that the result is simply a sum of the contributions of the constituent branes.

We investigated the behavior of the D2-branes wrapping the $\PP^1$s coming from the resolution of the singularity under the non-polynomial bulk perturbations and likewise calculated the corresponding superpotential.
It would be interesting to re-investigate this from a geometrical point of view. In a geometric setting, the problem is usually solved by going to an equivalent geometry, whereas this was not necessary in our approach. Hence, a geometric interpretation of our results would be of interest. For polynomial deformations, this was achieved in
 \cite{Baumgartl:2010ad}, where it was shown how Landau-Ginzburg superpotentials are related to period integrals. 
Once this is understood, geometric methods are very useful to understand the bulk deformations to higher order.

\medskip
\medskip
\medskip

\noindent
{\bf \large Acknowledgments:} We would like to thank A.~Collinucci, M.~Kay, P.~Mayr, M.~Soroush, and especially N.~Carqueville for discussions. This work was supported by a EURYI award of the European Science Foundation.

\appendix

\section{Gauge Equivalent Form of the 2d-Branch of Bound States}\label{sec:App_matrices}
\setcounter{equation}{0}
\numberwithin{equation}{section}

The $Q_2$-part of a matrix factorization $Q$ given by (\ref{eq:boundtensor})-(\ref{Tachyon}), can be represented in the following matrix form:
\begin{align}
 Q_2=\left(
\begin{array}{cccc}
 0 & 0 & E_A & 0\\
0& 0& f_0 & E_B \\
J_A & 0 & 0 & 0 \\
f_1 & J_B & 0 & 0
\end{array}
\right).
\end{align} 
Here, $f_0$, $f_1$, $E_{A/B}$, and $J_{A/B}$ are block matrices given by:
\begin{align}
 f_0=\left(
\begin{array}{cc}
  T_2 & T_3\\
  T_1 & -1
\end{array}
\right), \qquad
f_1=\left(
\begin{array}{cc}
 1 & T_3\\
 T_1 & -T_2
\end{array}
\right),
\end{align}
and
\begin{align}
 E_A=\left(
\begin{array}{cc}
  E_{2A} & E_{3A}\\
  J_{3A} & -J_{2A}
\end{array}
\right), \qquad
J_A=\left(
\begin{array}{cc}
 J_{2A} & E_{3A}\\
 J_{3A} & -E_{2A}
\end{array}
\right),
\end{align}
and analogously for $E_B$ and $J_B$. Using the transformation matrix:
\begin{align}
U=\left(
\begin{array}{cccccccc}
 0 & 0 & 1 & T_3 & 0 & 0 & 0 & 0 \\
 0 & 1 & 0 & -J_{2A} & 0 & 0 & 0 & 0 \\
 1 & T_3 & J_{2B} & E_{3A}-T_3 (J_{2A}-J_{2B}) & 0 & 0 & 0 & 0 \\
 0 & 0 & 0 & -1 & 0 & 0 & 0 & 0 \\
 0 & 0 & 0 & 0 & -T_1 & 1 & -J_{3B} & J_{2B} \\
 0 & 0 & 0 & 0 & 0 & 0 & 1 & 0 \\
 0 & 0 & 0 & 0 & 1 & 0 & -J_{2A} & 0 \\
 0 & 0 & 0 & 0 & 0 & 0 & -T_1 & 1
\end{array}
\right),
\end{align}
one can bring $Q_2$ to the form:
\begin{align}\label{eq:boundstatetransf}
\hat{Q}_2=U Q_2 U^{-1}=\left(
\begin{array}{cc}
 0 & \hat{E} \\
 \hat{J} & 0 
\end{array}
\right) \oplus \left(
\begin{array}{cc}
0 & W_2 \\
1 & 0
\end{array}
\right) \oplus \left(
\begin{array}{cc}
0 & W_2 \\
1 & 0
\end{array}
\right),
\end{align}
where
\begin{align}
\hat{E}=\left(
 \begin{array}{cc}
  T_2+ T_1 T_3 & E_{3A} - T_3 J_{2A}\\
  J_{3A} - T_1 J_{2A} & J_{2A} J_{2B}
 \end{array}
\right), \qquad 
\hat{J}=\left(
 \begin{array}{cc}
  -J_{2A} J_{2B} & E_{3A} - T_3 J_{2A}\\
  J_{3A} - T_1 J_{2A} & -T_2- T_1 T_3
 \end{array}
\right),
\end{align}
and $W_2$ is the factor associated to $Q_2$ in the tensor product decomposition, $W=W_1+W_2$.

\bibliographystyle{utcaps}
\bibliography{bound_v2}

\end{document}